\documentclass{article}%
\usepackage{amsfonts}
\usepackage{amsmath}
\usepackage{fullpage}
\usepackage{amssymb}
\usepackage{graphicx}%
\providecommand{\U}[1]{\protect\rule{.1in}{.1in}}
\newtheorem{theorem}{Theorem}

\newtheorem{lemma}[theorem]{Lemma}

\newenvironment{proof}[1][Proof]{\noindent\textbf{#1.} }{\ \rule{0.5em}{0.5em}}

\usepackage{color}

\begin{document}

\title{Joint source-channel coding for a quantum multiple access channel}
\author{Mark M. Wilde and Ivan Savov\\\textit{School of Computer Science, McGill University, Montreal, Quebec H3A
2A7}}
\maketitle

\begin{abstract}
Suppose that two senders each obtain one share of the output of a classical,
bivariate, correlated information source. They would like to transmit the
correlated source to a receiver using a quantum multiple access channel. In
prior work, Cover, El Gamal, and Salehi provided 
a combined source-channel coding strategy for a
classical multiple access channel 
which outperforms the simpler \textquotedblleft separation\textquotedblright\ strategy
where separate codebooks are used for the source coding 
and the channel coding tasks.
In the present paper, we prove that a coding
strategy similar to the Cover-El Gamal-Salehi strategy and a corresponding
quantum simultaneous decoder allow for the reliable transmission of a source
over a quantum multiple access channel, as long as a set of information
inequalities involving the Holevo quantity hold. 

\end{abstract}

Suppose that a correlated information source, embodied in many realizations of
two random variables $U$ and $V$ distributed independently and identically\ as
$p\left(  u,v\right)  $, is in the possession of two spatially separated
senders, such that one sender has $U$ and the other $V$. Suppose further that
a multiple access channel, modeled as a conditional probability distribution
$p\left(  y|x_{1},x_{2}\right)  $, connects these two senders to a single
receiver. 
When is
it possible for the senders to transmit the output of the correlated source
reliably over the multiple access channel? A \textit{separation} strategy
would have the senders first use a Slepian-Wolf compression code
\cite{SW73}\ to compress the source, followed by encoding the compressed bits
using a multiple access channel code \cite{L72,A74}. The receiver would first
decode the multiple access channel code and then decode the Slepian-Wolf
compression code in order to recover the output of the information source. This
strategy will work provided that the following information inequalities hold:%
\begin{align*}
H\left(  U|V\right)   &  \leq I\left(  X_{1};Y|X_{2}\right)  ,\\
H\left(  V|U\right)   &  \leq I\left(  X_{2};Y|X_{1}\right)  ,\\
H\left(  UV\right)   &  \leq I\left(  X_{1}X_{2};Y\right)  ,
\end{align*}
where the information quantities are with respect to a distribution of the
form $p\left(  u,v\right)  p\left(  x_{1}\right)  p\left(  x_{2}\right)
p\left(  y|x_{1},x_{2}\right)  $. 
The codewords for the multiple access channel
code in this scheme are generated independently according to the
product distribution $p\left(x_{1}\right)  p\left(  x_{2}\right)$.

Cover, El Gamal, and Salehi (CES) demonstrated the failure of a separation
strategy in the above scenario \cite{CES80}, in spite of a separation strategy
being optimal for the case of a single sender and single receiver 
\cite{bell1948shannon}. They found a simple example of a source with $p\left(
0,0\right)  =p\left(  0,1\right)  =p\left(  1,1\right)  =1/3$ and $p\left(
1,0\right)  =0$ and a channel $Y=X_{1}+X_{2}$ such that the above information
inequalities do not hold ($H\left(  UV\right)  =\log_{2}3=1.58$ while
$\max_{p\left(  x_{1}\right)  p\left(  x_{2}\right)  }I\left(  X_{1}%
X_{2};Y\right)  =1.5$), whereas the simple \textquotedblleft
joint\textquotedblright\ strategy of sending the source directly over the
channel succeeds perfectly (set $X_{1}=U$ and $X_{2}=V$ so that the receiver
can determine $U$ and $V$ from $Y$). More generally, the main result of their
paper is a joint source-channel coding strategy to allow for the senders to
transmit a source reliably over the channel, provided that the following
information inequalities hold:%
\begin{align*}
H\left(  U|V\right)   &  \leq I\left(  X_{1};Y|X_{2}VS\right)  ,\\
H\left(  V|U\right)   &  \leq I\left(  X_{2};Y|X_{1}US\right)  ,\\
H\left(  UV|W\right)   &  \leq I\left(  X_{1}X_{2};Y|WS\right)  ,\\
H\left(  UV\right)   &  \leq I\left(  X_{1}X_{2};Y\right)  ,
\end{align*}
where the variable $W$ is
defined to be the \textquotedblleft common part\textquotedblright\ of $U$ and
$V$ (to be defined later),
and the information quantities are with respect to a distribution $p\left(
u,v\right)  p\left(  s\right)  p\left(  x_{1}|u,s\right)  p\left(
x_{2}|v,s\right)  p\left(  y|x_{1},x_{2}\right)  $.
The advantage of the combined source-channel coding strategy
is that the codebooks for the multiple access channel
are now allowed to be correlated, because they are generated 
conditionally from the correlated source $p\left(u,v\right)$.


We can also study the combined source-channel coding problem
in the context of quantum information theory. 
Quantum channels are models of communication which aim to 
fully represent the quantum degrees of freedom associated with 
a given communication medium \cite{ieee1998holevo,SW97}.
Understanding quantum degrees of freedom is of practical importance in the context
of optical communications (for the case of bosonic channels).
For example, it is well known that a quantum strategy at the receiver's end of an optical channel 
can outperform a classical strategy such as homodyne, heterodyne, or direct detection
\cite{PhysRevLett.92.027902,Guh10a,GTW12,GW12}. These latter results suggest
that we should consider the extension 
of the classical information theoretic results to the quantum domain in order to study the source-channel problem.

The CES example above demonstrates that a separation strategy is not optimal
in general for the case of a quantum multiple access channel
\cite{winter2001capacity}, in contrast to a separation strategy being optimal
for the single-sender single-receiver quantum channel \cite{DHW11}. (This
result follows simply because every classical channel is a special type of
quantum channel.) For these reasons, we feel that it is important to extend the CES theorem to the domain of
quantum information theory.


The main accomplishment of the present paper is a generalization of the CES
theorem to the case of quantum multiple access channels with classical inputs
$x_{1}$ and $x_{2}$ and quantum outputs $\rho_{x_{1},x_{2}}$. Our main result
is the following theorem:

\begin{theorem}
\label{thm:main-theorem}A source $\left(  U^{n},V^{n}\right)  \sim
\prod\limits_{i=1}^{n}p\left(  u_{i},v_{i}\right)  $ can be sent with
arbitrarily small probability of error over a cq multiple access channel
$\left(  \mathcal{X}_{1}\times\mathcal{X}_{2},\rho_{x_{1},x_{2}}%
,\mathcal{H}^{B}\right)  $, with allowed codes $\left\{  x_{1}^{n}\left(
u^{n}\right)  ,x_{2}^{n}\left(  v^{n}\right)  \right\}  $ if there exist
probability mass functions $p\left(  s\right)  $, $p\left(  x_{1}|u,s\right)
$, $p\left(  x_{2}|v,s\right)  $, such that the following information
inequalities hold%
\begin{align*}
H\left(  U|V\right)   &  \leq I\left(  X_{1};B|X_{2}VS\right)  ,\\
H\left(  V|U\right)   &  \leq I\left(  X_{2};B|X_{1}US\right)  ,\\
H\left(  UV|W\right)   &  \leq I\left(  X_{1}X_{2};B|WS\right)  ,\\
H\left(  UV\right)   &  \leq I\left(  X_{1}X_{2};B\right)  .
\end{align*}
The above information quantities are calculated with respect to the following
classical-quantum state:%
\begin{equation}
\sum_{u,v,w,s,x_{1},x_{2}}p\left(  u,v,w\right)  p\left(  s\right)  p\left(
x_{1}|u,s\right)  p\left(  x_{2}|v,s\right)  \left\vert u,v,w,s,x_{1}%
,x_{2}\right\rangle \left\langle u,v,w,s,x_{1},x_{2}\right\vert ^{UVWSX_{1}%
X_{2}}\otimes\rho_{x_{1},x_{2}}^{B}.\label{eq:code-state}%
\end{equation}
The distribution $p\left(  u,v,w\right)  $ is defined as%
\[
p\left(  u,v,w\right)  =p\left(  u,v\right)  \delta_{f\left(  u\right)
,w}\delta_{g\left(  v\right)  ,w},
\]
so that the random variable $W=f\left(  U\right)  =g\left(  V\right)  $
represents the common part of $\left(  U,V\right)  $. The auxiliary random variable $S$ also
has the cardinality bound $|\mathcal{S}| \leq \min\{|\mathcal{X}|\cdot |\mathcal{Y}|, \dim(\mathcal{H}_B)^2  \}$
by using the same method in Appendix~A of Ref.~\cite{YHD2006}.
\end{theorem}

Our proof of the above theorem is significantly different from the way that
CES\ proved the classical version in Ref.~\cite{CES80} (and even different
from the later simplification in Ref.~\cite{AH83}). First, we require the use
of quantum typical projectors \cite{ieee1998holevo,SW97,W11} (reviewed in an
appendix), and furthermore, we exploit and extend several recent advances made in the
context of the quantum interference channel \cite{S11a,FHSSW11} (later applied
in other contexts such as the quantum broadcast channel~\cite{SW11}, the
quantum relay channel \cite{SWV12}, and an entanglement-assisted quantum
multiple access channel \cite{XW11}). Even if one were to consider the case of
our approach applied to a classical channel, our approach is different from
that in Refs.~\cite{CES80,AH83} (consider how our decomposition of the
\textquotedblleft error events\textquotedblright\ in
(\ref{eq:sum-decomposition}) differs from the error event decomposition in
Ref.~\cite{CES80}).

A quantum successive decoding strategy \cite{winter2001capacity,FHSSW11}%
\ cannot achieve the region given in Theorem~\ref{thm:main-theorem},
demonstrating a clear advantage of the quantum simultaneous decoding technique
\cite{S11a,FHSSW11}\ in the setting of the quantum multiple access channel. We
expect these simultaneous decoding techniques to extend to many other
scenarios in network quantum information theory (indeed consider those already
developed in Refs.~\cite{SW11,SWV12,XW11}), in spite of the current limitation
of the approach to decoding the transmissions of at most two senders.

In the next section, we describe the setting of the problem and define the
information processing task that we are considering. Section~\ref{sec:proof}%
\ provides a detailed proof of Theorem~\ref{thm:main-theorem},\ and we
conclude with some final remarks and suggestions for open problems in
Section~\ref{sec:conclusion}.

\section{Information Processing Task}

Suppose that two IID\ information sources $U_{1}$, $U_{2}$, \ldots, $U_{n}$
and $V_{1}$, $V_{2}$, \ldots, $V_{n}$ with distribution%
\[
p\left(  u^{n},v^{n}\right)  \equiv\prod\limits_{i=1}^{n}p\left(  u_{i}%
,v_{i}\right)
\]
are available to two senders, respectively.

Suppose that it is possible to arrange the matrix for the 
joint distribution $p\left(  u,v\right)  $ into the following block-diagonal form%
\[
\bigoplus\limits_{w=1}^{k}%
\begin{bmatrix}
p\left(  u_{1}^{\left(  w\right)  },v_{1}^{\left(  w\right)  }\right)   &
\cdots & p\left(  u_{1}^{\left(  w\right)  },v_{m_{w}}^{\left(  i\right)
}\right)  \\
\vdots & \ddots & \vdots\\
p\left(  u_{l_{w}}^{\left(  w\right)  },v_{1}^{\left(  w\right)  }\right)   &
\cdots & p\left(  u_{l_{w}}^{\left(  w\right)  },v_{m_{w}}^{\left(  w\right)
}\right)
\end{bmatrix},
\]
for some $w \in \left\{  1,\ldots,k\right\}$ and integers $l_w$ and $m_w$.
Then the common part \cite{CES80,el2010lecture}\ of $U$ and $V$ is the random
variable $W$ that is equal to $w$ if $\left(  u,v\right)  $ is in block
$w\in\left\{  1,\ldots,k\right\}  $. Observe that it is possible to determine
the common part from either $U$ or $V$ alone, simply by determining the block to
which a realization $u$ or $v$ belongs. We can define this common part by
finding the maximum integer $k$ such that there exist functions $f$ and $g$%
\begin{align*}
f &  :U\rightarrow\left\{  1,\ldots,k\right\}  ,\\
g &  :V\rightarrow\left\{  1,\ldots,k\right\}  ,
\end{align*}
with $\Pr\left\{  f\left(  U\right)  =w\right\}  >0$, $\Pr\left\{  g\left(
V\right)  =w\right\}  >0$, for $w\in\left\{  1,\ldots,k\right\}  $ with
$\Pr\left\{  f\left(  U\right)  =g\left(  V\right)  \right\}  =1$. We define
$W=f\left(  U\right)  =g\left(  V\right)  $. One can then write the
distribution of $U$, $V$, and $W$ as%
\[
p\left(  u,v,w\right)  =p\left(  u,v\right)  \delta_{f\left(  u\right)
,w}\delta_{g\left(  v\right)  ,w}.
\]

A combined code for the source and the quantum multiple access channel
consists of two encoding functions:%
\begin{align*}
x_{1}^{n} &  :\mathcal{U}^{n}\rightarrow\mathcal{X}_{1}^{n},\\
x_{2}^{n} &  :\mathcal{V}^{n}\rightarrow\mathcal{X}_{2}^{n},
\end{align*}
that assign codewords suitable for input to the quantum multiple access
channel for each of the possible outputs of the source. Upon input of these
codewords $x_{1}^{n}\left(  u^{n}\right)  $ and $x_{2}^{n}\left(
v^{n}\right)  $ to the quantum multiple access channel, the
quantum output for the receiver is%
\[
\rho_{x_{1}^{n}\left(  u^{n}\right)  ,x_{2}^{n}\left(  v^{n}\right)  }%
=\rho_{x_{11}\left(  u^{n}\right)  ,x_{21}\left(  v^{n}\right)  }\otimes
\cdots\otimes\rho_{x_{1n}\left(  u^{n}\right)  ,x_{2n}\left(  v^{n}\right)  }.
\]
The receiver performs a decoding positive operator-valued measure (POVM)
$\left\{  \Lambda_{u^{n},v^{n}}\right\}  $ whose elements $\Lambda
_{u^{n},v^{n}}$ are positive and form a resolution of the identity:%
\begin{align*}
\Lambda_{u^{n},v^{n}} &  \geq0,\\
\sum_{\left(  u^{n},v^{n}\right)  \in\mathcal{U}^{n}\times\mathcal{V}^{n}%
}\Lambda_{u^{n},v^{n}} &  =I^{\otimes n}.
\end{align*}
The end-to-end probability of error $P_{e}^{\left(  n\right)  }$ in this
protocol is the probability that the source is decoded incorrectly by the
receiver:%
\[
P_{e}^{\left(  n\right)  }\equiv\sum_{\left(  u^{n},v^{n}\right)
\in\mathcal{U}^{n}\times\mathcal{V}^{n}}p\left(  u^{n},v^{n}\right)
\text{Tr}\left\{  \left(  I-\Lambda_{u^{n},v^{n}}\right)  \rho_{x_{1}%
^{n}\left(  u^{n}\right)  ,x_{2}^{n}\left(  v^{n}\right)  }\right\}  .
\]

We say that it is possible to transmit the source reliably if there exists a
sequence of block codes $\left\{  x_{1}^{n}\left(  u^{n}\right)  ,x_{2}%
^{n}\left(  v^{n}\right)  \right\}  $ and a corresponding decoding
POVM\ $\left\{  \Lambda_{u^{n},v^{n}}\right\}  $ such that $P_{e}^{\left(
n\right)  }\rightarrow0$ as $n\rightarrow\infty$.

\section{Achievability Proof}

\label{sec:proof}This section provides a proof of Theorem~\ref{thm:main-theorem}, by
allowing for the generation of a code randomly and analyzing the expectation
of the error probability under this random choice. The main steps in the proof are
the random construction of the codebooks, the design of a decoding measurement for the receiver,
and an error analysis demonstrating that the inequalities in Theorem~\ref{thm:main-theorem} are a sufficient
set of conditions for transmitting the correlated source reliably.
The random construction of the codebooks is identical to that in the CES paper
\cite{CES80}, but the difference for the quantum case
lies in designing a collective decoding measurement for the receiver (this is always the
difference between any protocol for classical communication over a classical channel and one for
classical communication over a quantum channel). Our decoding POVM exploits and
extends the quantum simultaneous decoding techniques from Refs.~\cite{S11a,FHSSW11}, though
the situation considered here is rather different from the one considered there. First, we demand that the receiver
only decode source sequences that are jointly typical. It is natural that such an approach should perform
well asymptotically, given that nearly all of the source probability concentrates on this set.
Second, our decoding POVM has several conditionally typical
projectors ``sandwiched'' together in a very particular order, such that our error analysis can proceed
in bounding errors in terms of the information quantities appearing in Theorem~\ref{thm:main-theorem}. The
decoding measurement is a square-root measurement \cite{ieee1998holevo,SW97}, though nothing would prevent
us from employing a sequential, simultaneous decoder of the flavor in Ref.~\cite{S11a}.
Our error analysis carefully employs indicator functions over the jointly typical set of source sequences
(since we are decoding only over this set), and this approach allows
us to introduce distributions that are needed for taking expectations of quantum codeword states.
Finally, we decompose error terms in a way different from that in the CES paper \cite{CES80}, and in our opinion,
this error
decomposition appears to be a bit more natural than the error event decomposition appearing there.

\bigskip

\begin{proof}
We split the proof into several stages.\bigskip

\textbf{Code Generation.} For every $w^{n}\in\mathcal{W}^{n}$, independently
generate one $s^{n}$ sequence according to%
\[
p\left(  s^{n}\right)  =\prod\limits_{i=1}^{n}p\left(  s_{i}\right)  .
\]
Index them as $s^{n}\left(  w^{n}\right)  $ for all $w^{n}\in\mathcal{W}^{n}$.

For every $u^{n}\in\mathcal{U}^{n}$, compute the corresponding $w^{n}%
=f^{n}\left(  u^{n}\right)  \equiv f\left(  u_{1}\right)  \cdots f\left(
u_{n}\right)  $ and independently generate one $x_{1}^{n}$ sequence according
to%
\[
p\left(  x_{1}^{n}|u^{n},s^{n}\left(  w^{n}\right)  \right)  =\prod
\limits_{i=1}^{n}p\left(  x_{1i}|u_{i},s_{i}\left(  w^{n}\right)  \right)  .
\]
Index these $x_{1}^{n}$ sequences according to $x_{1}^{n}\left(  u^{n}%
,s^{n}\left(  w^{n}\right)  \right)  $.

Similarly, for every $v^{n}\in\mathcal{V}^{n}$, compute the corresponding
$w^{n}=g^{n}\left(  v^{n}\right)  \equiv g\left(  v_{1}\right)  \cdots
g\left(  v_{n}\right)  $ and independently generate one $x_{2}^{n}$ sequence
according to%
\[
p\left(  x_{2}^{n}|v^{n},s^{n}\left(  w^{n}\right)  \right)  =\prod
\limits_{i=1}^{n}p\left(  x_{2i}|v_{i},s_{i}\left(  w^{n}\right)  \right)  .
\]
Index these $x_{2}^{n}$ sequences according to $x_{2}^{n}\left(  v^{n}%
,s^{n}\left(  w^{n}\right)  \right)  $.\bigskip

\textbf{Encoding}. Upon observing the output $u^{n}$\ of the source, Sender~1
computes $w^{n}=f^{n}\left(  u^{n}\right)  $ and transmits $x_{1}^{n}\left(
u^{n},s^{n}\left(  w^{n}\right)  \right)  $. Similarly, upon observing the
output $v^{n}$\ of the source, Sender~2 computes $w^{n}=g^{n}\left(
v^{n}\right)  $ and transmits $x_{2}^{n}\left(  v^{n},s^{n}\left(
w^{n}\right)  \right)  $.\bigskip

\textbf{Decoding POVM}. We now need to determine a decoding POVM\ that the
receiver can perform to detect what codewords the senders transmitted. In this
vein, we need to consider several typical projectors (see
Appendix~\ref{sec:typ-review}) corresponding to several reduced states of
(\ref{eq:code-state}). By considering the following information equalities, we
can determine which reduced states we will require:%
\begin{align*}
I\left(  X_{1};B|X_{2}VS\right)   &  =H\left(  B|X_{2}VS\right)  -H\left(
B|X_{1}X_{2}VS\right)  \\
&  =H\left(  B|X_{2}VS\right)  -H\left(  B|X_{1}X_{2}\right)  ,\\
I\left(  X_{2};B|X_{1}US\right)   &  =H\left(  B|X_{1}US\right)  -H\left(
B|X_{1}X_{2}US\right)  \\
&  =H\left(  B|X_{1}US\right)  -H\left(  B|X_{1}X_{2}\right)  ,\\
I\left(  X_{1}X_{2};B|WS\right)   &  =H\left(  B|WS\right)  -H\left(
B|X_{1}X_{2}WS\right)  \\
&  =H\left(  B|WS\right)  -H\left(  B|X_{1}X_{2}\right)  ,\\
I\left(  X_{1}X_{2};B\right)   &  =H\left(  B\right)  -H\left(  B|X_{1}%
X_{2}\right)  .
\end{align*}
Thus, it is clear that five different reduced states corresponding to the
entropies $H\left(  B|X_{2}VS\right)  $, $H\left(  B|X_{1}US\right)  $,
$H\left(  B|WS\right)  $, $H\left(  B\right)  $, and $H\left(  B|X_{1}%
X_{2}\right)  $ are important.

First, we consider the reduced state on $X_{2}VSB$:%
\[
\sum_{v,s,x_{2}}p\left(  v\right)  p\left(  s\right)  p\left(  x_{2}%
|v,s\right)  \ \left\vert v,s,x_{2}\right\rangle \left\langle v,s,x_{2}%
\right\vert ^{VSX_{2}}\otimes\sigma_{v,s,x_{2}}^{B},
\]
where we define the states $\sigma_{v,s,x_{2}}^{B}$ as%
\begin{align}
\sigma_{v,s,x_{2}}^{B} &  \equiv\sum_{u,w,x_{1}}p\left(  u,w|v\right)
p\left(  x_{1}|u,s\right)  \rho_{x_{1},x_{2}}^{B}\label{eq:sigma-state}\\
&  =\sum_{u,x_{1}}p\left(  u|v\right)  p\left(  x_{1}|u,s\right)  \rho
_{x_{1},x_{2}}^{B}.\nonumber
\end{align}
The second equality follows from (\ref{eq:source-1}). 
Corresponding to a tensor product $\sigma_{v^{n},s^{n},x_{2}^{n}}^{B^{n}%
}\equiv\sigma_{v_{1},s_{1},x_{21}}^{B_{1}}\otimes\cdots\otimes\sigma
_{v_{n},s_{n},x_{2n}}^{B_{n}}$ of these states is a conditionally typical
projector $\Pi_{\sigma_{v^{n},s^{n},x_{2}^{n}}}$ which we abbreviate as
$\Pi_{x_{2}^{n}\left(  v^{n},s^{n}\right)  }$. The rank of each of these
conditionally typical projectors is approximately equal to $2^{nH\left(
B|X_{2}VS\right)  }$.
Let $\rho_{x_{2},s}^{\left(  u\right)  }$ denote the following state:%
\begin{equation}
\rho_{x_{2},s}^{\left(  u\right)  }\equiv\sum_{x_{1}}p\left(  x_{1}%
|u,s\right)  \rho_{x_{1},x_{2}}^{B}.\label{eq:u-state}%
\end{equation}

Next, we consider the reduced state on $X_{1}USB$:%
\[
\sum_{u,s,x_{1}}p\left(  u\right)  p\left(  s\right)  p\left(  x_{1}%
|u,s\right)  \ \left\vert u,s,x_{1}\right\rangle \left\langle u,s,x_{1}%
\right\vert ^{USX_{1}}\otimes\omega_{u,s,x_{1}}^{B},
\]
where we define the states $\omega_{u,s,x_{1}}^{B}
=\sum_{v,x_{2}}p\left(  v|u\right)  p\left(  x_{2}|v,s\right)  \rho_{x_{1},x_{2}}^{B}
$.
Corresponding to a tensor product $\omega_{u^{n},s^{n},x_{1}^{n}}^{B^{n}}$ of
these states is a conditionally typical projector $\Pi_{\omega_{u^{n}%
,s^{n},x_{1}^{n}}}$ which we abbreviate as $\Pi_{x_{1}^{n}\left(  u^{n}%
,s^{n}\right)  }$. The rank of each of these conditionally typical projectors
is approximately equal to $2^{nH\left(  B|X_{1}US\right)  }$.
Let $\rho_{x_{1}%
,s}^{\left(  v\right)  }$ denote the following state:%
\begin{equation}
\rho_{x_{1},s}^{\left(  v\right)  }\equiv\sum_{x_{2}}p\left(  x_{2}%
|v,s\right)  \rho_{x_{1},x_{2}}^{B}. \label{eq:v-state}%
\end{equation}

Third, we consider the reduced state on $WSB$:%
\[
\sum_{w,s}p\left(  w\right)  p\left(  s\right)  \left\vert w,s\right\rangle
\left\langle w,s\right\vert ^{WS}\otimes\tau_{w,s}^{B},
\]
where the states $\tau_{w,s}^{B}$\ are defined as%
\begin{equation}
\tau_{w,s}^{B}\equiv\sum_{x_{1},x_{2},u,v}p\left(  u,v|w\right)  p\left(
x_{1}|u,s\right)  p\left(  x_{2}|v,s\right)  \rho_{x_{1},x_{2}}^{B}%
.\label{eq:tau-state}%
\end{equation}
Corresponding to a tensor product $\tau_{w^{n},s^{n}}^{B^{n}}$ of these states
is a conditionally typical projector $\Pi_{\tau_{w^{n},s^{n}}}$ which we
abbreviate as $\Pi_{s^{n}\left(  w^{n}\right)  }$. The rank of each of these
conditionally typical projectors is approximately equal to $2^{nH\left(
B|WS\right)  }$.

Fourth, we have the state $\overline{\rho}^{B}$ obtained by tracing over every
system except $B$:%
\[
\overline{\rho}^{B}\equiv\sum_{u,v,w,s,x_{1},x_{2}}p\left(  u,v,w\right)
p\left(  s\right)  p\left(  x_{1}|u,s\right)  p\left(  x_{2}|v,s\right)
\rho_{x_{1},x_{2}}^{B}.
\]
To a tensor power $\left(  \overline{\rho}^{B}\right)  ^{\otimes n}$ of this
state corresponds a typical projector $\Pi_{\overline{\rho}^{\otimes n}}$,
which we abbreviate as just $\Pi$. The rank of this projector is approximately
equal to $2^{nH\left(  B\right)  }$.

Finally, we define the tensor product states $\rho_{x_{1}^{n},x_{2}^{n}%
}^{B^{n}}$, which correspond to the outputs of the channel when $x_{1}^{n}$
and $x_{2}^{n}$ are input. To these states corresponds a conditionally typical
projector $\Pi_{\rho_{x_{1}^{n},x_{2}^{n}}}$, which we abbreviate as
$\Pi_{x_{1}^{n},x_{2}^{n}}$. Its rank is approximately equal to $2^{nH\left(
B|X_{1}X_{2}\right)  }$.

We can now construct the decoding POVM\ for the receiver. For every $\left(
u^{n},v^{n},w^{n}\right)  \in\mathcal{T}^{n}$ (with $\mathcal{T}^{n}$ the
joint weakly typical set defined in
Appendix~\ref{sec:indicator-function-trick}), consider the following
operators:%
\begin{equation}
\Gamma_{u^{n},v^{n},w^{n}}\equiv\Pi\ \Pi_{s^{n}\left(  w^{n}\right)  }%
\ \Pi_{x_{1}^{n}\left(  u^{n},s^{n}\right)  }\ \Pi_{x_{1}^{n}\left(
u^{n},s^{n}\right)  ,x_{2}^{n}\left(  v^{n},s^{n}\right)  }\ \Pi_{x_{1}%
^{n}\left(  u^{n},s^{n}\right)  }\ \Pi_{s^{n}\left(  w^{n}\right)  }\ \Pi.
\label{eq:gamma-def}%
\end{equation}
Note that we define these operators only over $\left(  u^{n},v^{n}%
,w^{n}\right)  \in\mathcal{T}^{n}$ and set them equal to zero otherwise. The
measurement for the receiver is a \textquotedblleft
square-root\textquotedblright\ measurement of the following form:%
\[
\Lambda_{u^{n},v^{n},w^{n}}\equiv\left(  \sum_{\left(  u^{\prime n},v^{\prime
n},w^{\prime n}\right)  \in\mathcal{T}^{n}}\Gamma_{u^{\prime n},v^{\prime
n},w^{\prime n}}\right)  ^{-1/2}\Gamma_{u^{n},v^{n},w^{n}}\left(
\sum_{\left(  u^{\prime n},v^{\prime n},w^{\prime n}\right)  \in
\mathcal{T}^{n}}\Gamma_{u^{\prime n},v^{\prime n},w^{\prime n}}\right)
^{-1/2}.
\]
\bigskip

\textbf{Error Analysis}. The decoding error is defined as follows:%
\[
\sum_{\left(  u^{n},v^{n}\right)  \in\mathcal{U}^{n}\times\mathcal{V}^{n}%
}p\left(  u^{n},v^{n}\right)  \Pr\left\{  \text{error made at decoder
}|\ \left(  u^{n},v^{n}\right)  \text{ is the output of the source}\right\}  .
\]
For our case, this reduces to%
\begin{align*}
&\!\!\!\!\!\!\!\!\!\!  \sum_{u^{n},v^{n}}p\left(  u^{n},v^{n}\right)  \text{Tr}\left\{  \left(
I-\Lambda_{u^{n},v^{n},f^{n}\left(  u^{n}\right)  }\right)  \rho_{x_{1}%
^{n}\left(  u^{n},s^{n}\right)  ,x_{2}^{n}\left(  v^{n},s^{n}\right)
}\right\}  \\
&  =\sum_{u^{n},v^{n},w^{n}}p\left(  u^{n},v^{n},w^{n}\right)  \text{Tr}%
\left\{  \left(  I-\Lambda_{u^{n},v^{n},w^{n}}\right)  \rho_{x_{1}^{n}\left(
u^{n},s^{n}\right)  ,x_{2}^{n}\left(  v^{n},s^{n}\right)  }\right\}  .
\end{align*}
We also allow for an expectation over the random choice of code. We will
introduce this when needed in order to simplify the error analysis (at the
end, we will determine that the expectation is not necessary---that there
exists some code by a  derandomization argument):%
\[
\sum_{u^{n},v^{n},w^{n}}p\left(  u^{n},v^{n},w^{n}\right)  \mathbb{E}_{S^{n}%
}\ \mathbb{E}_{X_{1}^{n}|u^{n}S^{n},X_{2}^{n}|v^{n}S^{n}}\left\{
\text{Tr}\left\{  \left(  I-\Lambda_{u^{n},v^{n},w^{n}}\right)  \rho
_{X_{1}^{n}\left(  u^{n},S^{n}\right)  ,X_{2}^{n}\left(  v^{n},S^{n}\right)
}\right\}  \right\}  .
\]
Our first \textquotedblleft move\textquotedblright\ is to smooth the state by
the projector $\Pi_{x_{2}^{n}\left(  v^{n},s^{n}\right)  }$ (this is
effectively \textquotedblleft smoothing\textquotedblright\ the channel). That
is, we can bound the error from above as%
\begin{multline}
\sum_{u^{n},v^{n},w^{n}}p\left(  u^{n},v^{n},w^{n}\right)  \text{Tr}\left\{
\left(  I-\Lambda_{u^{n},v^{n},w^{n}}\right)  \ \Pi_{X_{2}^{n}\left(
v^{n},S^{n}\right)  }\ \rho_{X_{1}^{n}\left(  u^{n},S^{n}\right)  ,X_{2}%
^{n}\left(  v^{n},S^{n}\right)  }\ \Pi_{X_{2}^{n}\left(  v^{n},S^{n}\right)
}\right\}  \label{eq:first-move}\\
+\sum_{u^{n},v^{n},w^{n}}p\left(  u^{n},v^{n},w^{n}\right)  \left\Vert
\rho_{X_{1}^{n}\left(  u^{n},S^{n}\right)  ,X_{2}^{n}\left(  v^{n}%
,S^{n}\right)  }-\Pi_{X_{2}^{n}\left(  v^{n},S^{n}\right)  }\ \rho_{X_{1}%
^{n}\left(  u^{n},S^{n}\right)  ,X_{2}^{n}\left(  v^{n},S^{n}\right)  }%
\ \Pi_{X_{2}^{n}\left(  v^{n},S^{n}\right)  }\right\Vert _{1}.
\end{multline}
By introducing the expectation over the random choice of code $\mathbb{E}%
_{S^{n}}\ \mathbb{E}_{X_{1}^{n}|u^{n}S^{n},X_{2}^{n}|v^{n}S^{n}}$, we can
bound the second term from above by $2\sqrt{\epsilon}$, by applying the Gentle
Operator Lemma for ensembles (Appendix~\ref{sec:useful-lemmas}) and the
properties of quantum typicality:%
\begin{multline*}
\sum_{u^{n},v^{n},w^{n}}p\left(  u^{n},v^{n},w^{n}\right)  \times\\
\mathbb{E}_{S^{n}}\ \mathbb{E}_{X_{1}^{n}|u^{n}S^{n},X_{2}^{n}|v^{n}S^{n}%
}\left\{  \left\Vert \rho_{X_{1}^{n}\left(  u^{n},S^{n}\right)  ,X_{2}%
^{n}\left(  v^{n},S^{n}\right)  }-\Pi_{X_{2}^{n}\left(  v^{n},S^{n}\right)
}\ \rho_{X_{1}^{n}\left(  u^{n},S^{n}\right)  ,X_{2}^{n}\left(  v^{n}%
,S^{n}\right)  }\ \Pi_{X_{2}^{n}\left(  v^{n},S^{n}\right)  }\right\Vert
_{1}\right\}  \leq2\sqrt{\epsilon}%
\end{multline*}

We now consider the first term in (\ref{eq:first-move}) and make the
abbreviation%
\[
\rho_{X_{1}^{n}\left(  u^{n},S^{n}\right)  ,X_{2}^{n}\left(  v^{n}%
,S^{n}\right)  }^{\prime}\equiv\Pi_{X_{2}^{n}\left(  v^{n},S^{n}\right)
}\ \rho_{X_{1}^{n}\left(  u^{n},S^{n}\right)  ,X_{2}^{n}\left(  v^{n}%
,S^{n}\right)  }\ \Pi_{X_{2}^{n}\left(  v^{n},S^{n}\right)  }.
\]
This term naturally splits up into two parts according to the typicality
condition $\left(  u^{n},v^{n},w^{n}\right)  \in\mathcal{T}^{n}$:%
\begin{multline}
\sum_{u^{n},v^{n},w^{n}}p\left(  u^{n},v^{n},w^{n}\right)  \left\{
\text{Tr}\left\{  \left(  I-\Lambda_{u^{n},v^{n},w^{n}}\right)  \ \rho
_{X_{1}^{n}\left(  u^{n},S^{n}\right)  ,X_{2}^{n}\left(  v^{n},S^{n}\right)
}^{\prime}\right\}  \right\}  \left[  \mathcal{I}\left(  \left(  u^{n}%
,v^{n},w^{n}\right)  \in\mathcal{T}^{n}\right)  +\mathcal{I}\left(  \left(
u^{n},v^{n},w^{n}\right)  \notin\mathcal{T}^{n}\right)  \right]  \\
=\sum_{u^{n},v^{n},w^{n}}p\left(  u^{n},v^{n},w^{n}\right)  \left\{
\text{Tr}\left\{  \left(  I-\Lambda_{u^{n},v^{n},w^{n}}\right)  \ \rho
_{X_{1}^{n}\left(  u^{n},S^{n}\right)  ,X_{2}^{n}\left(  v^{n},S^{n}\right)
}^{\prime}\right\}  \right\}  \mathcal{I}\left(  \left(  u^{n},v^{n}%
,w^{n}\right)  \in\mathcal{T}^{n}\right)  +\\
\sum_{u^{n},v^{n},w^{n}}p\left(  u^{n},v^{n},w^{n}\right)  \left\{
\text{Tr}\left\{  \rho_{X_{1}^{n}\left(  u^{n},S^{n}\right)  ,X_{2}^{n}\left(
v^{n},S^{n}\right)  }^{\prime}\right\}  \right\}  \mathcal{I}\left(  \left(
u^{n},v^{n},w^{n}\right)  \notin\mathcal{T}^{n}\right)
\label{eq:term-to-double}%
\end{multline}
where in the last line we used that $\Lambda_{u^{n},v^{n},w^{n}}=0$ whenever
$\left(  u^{n},v^{n},w^{n}\right)  \notin\mathcal{T}^{n}$. (We have also
introduced an indicator function $\mathcal{I}\left(  \left(  u^{n},v^{n}%
,w^{n}\right)  \in\mathcal{T}^{n}\right)  $ which is equal to one if $\left(
u^{n},v^{n},w^{n}\right)  \in\mathcal{T}^{n}$ equal to zero otherwise.) We now
apply the well-known Hayashi-Nagaoka operator inequality (Lemma 2 of
Ref.~\cite{hayashi2003general})\ to the term in the middle line, giving the
upper bound%
\begin{multline*}
2\sum_{u^{n},v^{n},w^{n}}p\left(  u^{n},v^{n},w^{n}\right)  \left\{
\text{Tr}\left\{  \left(  I-\Gamma_{u^{n},v^{n},w^{n}}\right)  \ \rho
_{X_{1}^{n}\left(  u^{n},S^{n}\right)  ,X_{2}^{n}\left(  v^{n},S^{n}\right)
}^{\prime}\right\}  \right\}  \mathcal{I}\left(  \left(  u^{n},v^{n}%
,w^{n}\right)  \in\mathcal{T}^{n}\right)  +\\
4\sum_{u^{n},v^{n},w^{n}}p\left(  u^{n},v^{n},w^{n}\right)  \sum_{\left(
u^{\prime n},v^{\prime n},w^{\prime n}\right)  \neq\left(  u^{n},v^{n}%
,w^{n}\right)  }\left\{  \text{Tr}\left\{  \Gamma_{u^{\prime n},v^{\prime
n},w^{\prime n}}\ \rho_{X_{1}^{n}\left(  u^{n},S^{n}\right)  ,X_{2}^{n}\left(
v^{n},S^{n}\right)  }^{\prime}\right\}  \right\}  \times\\
\mathcal{I}\left(  \left(  u^{n},v^{n},w^{n}\right)  \in\mathcal{T}%
^{n}\right)  \mathcal{I}\left(  \left(  u^{\prime n},v^{\prime n},w^{\prime
n}\right)  \in\mathcal{T}^{n}\right)
\end{multline*}
Doubling the term in the last line of (\ref{eq:term-to-double}) and adding it
to the first above gives the following upper bound%
\begin{multline*}
2\sum_{u^{n},v^{n},w^{n}}p\left(  u^{n},v^{n},w^{n}\right)  \left\{
\text{Tr}\left\{  \left(  I-\Gamma_{u^{n},v^{n},w^{n}}\right)  \ \rho
_{X_{1}^{n}\left(  u^{n},S^{n}\right)  ,X_{2}^{n}\left(  v^{n},S^{n}\right)
}^{\prime}\right\}  \right\}  +\\
4\sum_{u^{n},v^{n},w^{n}}p\left(  u^{n},v^{n},w^{n}\right)  \sum_{\left(
u^{\prime n},v^{\prime n},w^{\prime n}\right)  \neq\left(  u^{n},v^{n}%
,w^{n}\right)  }\left\{  \text{Tr}\left\{  \Gamma_{u^{\prime n},v^{\prime
n},w^{\prime n}}\ \rho_{X_{1}^{n}\left(  u^{n},S^{n}\right)  ,X_{2}^{n}\left(
v^{n},S^{n}\right)  }^{\prime}\right\}  \right\}  \times\\
\mathcal{I}\left(  \left(  u^{n},v^{n},w^{n}\right)  \in\mathcal{T}%
^{n}\right)  \mathcal{I}\left(  \left(  u^{\prime n},v^{\prime n},w^{\prime
n}\right)  \in\mathcal{T}^{n}\right)
\end{multline*}
We bound the first term, by introducing the expectation $\mathbb{E}_{S^{n}%
}\ \mathbb{E}_{X_{1}^{n}|u^{n}S^{n},X_{2}^{n}|v^{n}S^{n}}$ and considering
that%
\begin{align*}
&  \sum_{u^{n},v^{n},w^{n}}p\left(  u^{n},v^{n},w^{n}\right)  \mathbb{E}%
_{S^{n}}\ \mathbb{E}_{X_{1}^{n}|u^{n}S^{n},X_{2}^{n}|v^{n}S^{n}}\left\{
\text{Tr}\left\{  \Gamma_{u^{n},v^{n},w^{n}}\ \rho_{X_{1}^{n}\left(
u^{n},S^{n}\right)  ,X_{2}^{n}\left(  v^{n},S^{n}\right)  }^{\prime}\right\}
\right\}  \\
&  =\mathbb{E}_{U^{n}V^{n}W^{n}S^{n}X_{1}^{n}X_{2}^{n}}\ \left\{
\text{Tr}\left\{  \Gamma_{U^{n},V^{n},X^{n}}\ \rho_{X_{1}^{n}\left(
U^{n},S^{n}\right)  ,X_{2}^{n}\left(  V^{n},S^{n}\right)  }^{\prime}\right\}
\right\}  \\
&  =\mathbb{E}_{U^{n}V^{n}W^{n}S^{n}X_{1}^{n}X_{2}^{n}}\ \left\{
\text{Tr}\left\{
\begin{array}
[c]{c}%
\Pi\ \Pi_{S^{n}\left(  W^{n}\right)  }\ \Pi_{X_{1}^{n}\left(  U^{n}%
,S^{n}\right)  }\ \Pi_{X_{1}^{n}\left(  U^{n},S^{n}\right)  ,X_{2}^{n}\left(
V^{n},S^{n}\right)  }\ \Pi_{X_{1}^{n}\left(  U^{n},S^{n}\right)  }\ \Pi
_{S^{n}\left(  W^{n}\right)  }\ \times\\
\Pi\ \Pi_{X_{2}^{n}\left(  V^{n},S^{n}\right)  }\ \rho_{X_{1}^{n}\left(
U^{n},S^{n}\right)  ,X_{2}^{n}\left(  V^{n},S^{n}\right)  }\ \Pi_{X_{2}%
^{n}\left(  V^{n},S^{n}\right)  }%
\end{array}
\right\}  \right\}  \\
&  \geq\mathbb{E}_{U^{n}V^{n}W^{n}S^{n}X_{1}^{n}X_{2}^{n}}\ \left\{
\text{Tr}\left\{  \Pi_{X_{1}^{n}\left(  U^{n},S^{n}\right)  ,X_{2}^{n}\left(
V^{n},S^{n}\right)  }\ \rho_{X_{1}^{n}\left(  U^{n},S^{n}\right)  ,X_{2}%
^{n}\left(  V^{n},S^{n}\right)  }\right\}  \right\}  \\
&  \ \ \ \ \ -\mathbb{E}_{U^{n}V^{n}W^{n}S^{n}X_{1}^{n}X_{2}^{n}}\ \left\{
\left\Vert \rho_{X_{1}^{n}\left(  U^{n},S^{n}\right)  ,X_{2}^{n}\left(
V^{n},S^{n}\right)  }-\Pi_{X_{2}^{n}\left(  V^{n},S^{n}\right)  }\ \rho
_{X_{1}^{n}\left(  U^{n},S^{n}\right)  ,X_{2}^{n}\left(  V^{n},S^{n}\right)
}\ \Pi_{X_{2}^{n}\left(  V^{n},S^{n}\right)  }\right\Vert _{1}\right\}  \\
&  \ \ \ \ \ -\mathbb{E}_{U^{n}V^{n}W^{n}S^{n}X_{1}^{n}X_{2}^{n}}\ \left\{
\left\Vert \rho_{X_{1}^{n}\left(  U^{n},S^{n}\right)  ,X_{2}^{n}\left(
V^{n},S^{n}\right)  }-\Pi\ \rho_{X_{1}^{n}\left(  U^{n},S^{n}\right)
,X_{2}^{n}\left(  V^{n},S^{n}\right)  }\ \Pi\right\Vert _{1}\right\}  \\
&  \ \ \ \ \ -\mathbb{E}_{U^{n}V^{n}W^{n}S^{n}X_{1}^{n}X_{2}^{n}}\ \left\{
\left\Vert \rho_{X_{1}^{n}\left(  U^{n},S^{n}\right)  ,X_{2}^{n}\left(
V^{n},S^{n}\right)  }-\Pi_{S^{n}\left(  W^{n}\right)  }\ \rho_{X_{1}%
^{n}\left(  U^{n},S^{n}\right)  ,X_{2}^{n}\left(  V^{n},S^{n}\right)  }%
\ \Pi_{S^{n}\left(  W^{n}\right)  }\right\Vert _{1}\right\}  \\
&  \ \ \ \ \ -\mathbb{E}_{U^{n}V^{n}W^{n}S^{n}X_{1}^{n}X_{2}^{n}}\ \left\{
\left\Vert \rho_{X_{1}^{n}\left(  U^{n},S^{n}\right)  ,X_{2}^{n}\left(
V^{n},S^{n}\right)  }-\Pi_{X_{1}^{n}\left(  U^{n},S^{n}\right)  }\ \rho
_{X_{1}^{n}\left(  U^{n},S^{n}\right)  ,X_{2}^{n}\left(  V^{n},S^{n}\right)
}\ \Pi_{X_{1}^{n}\left(  U^{n},S^{n}\right)  }\right\Vert _{1}\right\}  \\
&  \geq1-\epsilon-8\sqrt{\epsilon},
\end{align*}
where we have used the trace inequality Tr$\left\{  \Lambda\rho\right\}
\geq\ $Tr$\left\{  \Lambda\sigma\right\}  -\left\Vert \rho-\sigma\right\Vert
_{1}$, properties of quantum typicality, and the Gentle Operator Lemma for
ensembles. Thus, we obtain the following upper bound on the error:%
\begin{multline*}
2\left(  \epsilon+8\sqrt{\epsilon}\right)  +4\sum_{u^{n},v^{n},w^{n}}p\left(
u^{n},v^{n},w^{n}\right)  \times\\
\sum_{\left(  u^{\prime n},v^{\prime n},w^{\prime n}\right)  \neq\left(
u^{n},v^{n},w^{n}\right)  }\left\{  \text{Tr}\left\{  \Gamma_{u^{\prime
n},v^{\prime n},w^{\prime n}}\ \rho_{X_{1}^{n}\left(  u^{n},S^{n}\right)
,X_{2}^{n}\left(  v^{n},S^{n}\right)  }^{\prime}\right\}  \right\}
\mathcal{I}\left(  \left(  u^{n},v^{n},w^{n}\right)  \in\mathcal{T}%
^{n}\right)  \mathcal{I}\left(  \left(  u^{\prime n},v^{\prime n},w^{\prime
n}\right)  \in\mathcal{T}^{n}\right)  .
\end{multline*}

The sum $\sum_{\left(  u^{\prime n},v^{\prime n},w^{\prime n}\right)
\neq\left(  u^{n},v^{n},w^{n}\right)  }\left(  \cdot\right)  $ can split up in
the following seven different ways:%
\[
\sum_{\left(  u^{\prime n},v^{\prime n},w^{\prime n}\right)  \neq\left(
u^{n},v^{n},w^{n}\right)  }\left(  \cdot\right)  =\sum_{\substack{u^{\prime
n}\neq u^{n},\\v^{\prime n}=v^{n},\\w^{\prime n}=w^{n}}}\left(  \cdot\right)
+\sum_{\substack{u^{\prime n}=u^{n},\\v^{\prime n}\neq v^{n},\\w^{\prime
n}=w^{n}}}\left(  \cdot\right)  +\sum_{\substack{u^{\prime n}=u^{n}%
,\\v^{\prime n}=v^{n},\\w^{\prime n}\neq w^{n}}}\left(  \cdot\right)
+\sum_{\substack{u^{\prime n}\neq u^{n},\\v^{\prime n}\neq v^{n},\\w^{\prime
n}=w^{n}}}\left(  \cdot\right)  +\sum_{\substack{u^{\prime n}\neq
u^{n},\\v^{\prime n}=v^{n},\\w^{\prime n}\neq w^{n}}}\left(  \cdot\right)
+\sum_{\substack{u^{\prime n}=u^{n},\\v^{\prime n}\neq v^{n},\\w^{\prime
n}\neq w^{n}}}\left(  \cdot\right)  +\sum_{\substack{u^{\prime n}\neq
u^{n},\\v^{\prime n}\neq v^{n},\\w^{\prime n}\neq w^{n}}}\left(  \cdot\right)
.
\]
Though, considering that $u^{\prime n}=u^{n}$ implies that $f^{n}\left(
u^{\prime n}\right)  =f^{n}\left(  u^{n}\right)  $ and thus that $w^{\prime
n}=w^{n}$, the terms in the following sums do not occur:%
\[
\sum_{\substack{u^{\prime n}=u^{n},\\v^{\prime n}=v^{n},\\w^{\prime n}\neq
w^{n}}}\left(  \cdot\right)  +\sum_{\substack{u^{\prime n}=u^{n},\\v^{\prime
n}\neq v^{n},\\w^{\prime n}\neq w^{n}}}\left(  \cdot\right)  .
\]
Similarly, we have that $v^{\prime n}=v^{n}$ implies that $g^{n}\left(
v^{\prime n}\right)  =g^{n}\left(  v^{n}\right)  $ and thus that $w^{\prime
n}=w^{n}$, and so the terms in the following sum do not occur:%
\[
\sum_{\substack{u^{\prime n}\neq u^{n},\\v^{\prime n}=v^{n},\\w^{\prime n}\neq
w^{n}}}\left(  \cdot\right)  .
\]
So this leaves us with the following decomposition of the overall sum into
four different sums:%
\begin{equation}
\sum_{\left(  u^{\prime n},v^{\prime n},w^{\prime n}\right)  \neq\left(
u^{n},v^{n},w^{n}\right)  }\left(  \cdot\right)  =\underbrace{\sum
_{\substack{u^{\prime n}\neq u^{n},\\v^{\prime n}=v^{n},\\w^{\prime n}=w^{n}%
}}\left(  \cdot\right)  }_{\alpha}+\underbrace{\sum_{\substack{u^{\prime
n}=u^{n},\\v^{\prime n}\neq v^{n},\\w^{\prime n}=w^{n}}}\left(  \cdot\right)
}_{\beta}+\underbrace{\sum_{\substack{u^{\prime n}\neq u^{n},\\v^{\prime
n}\neq v^{n},\\w^{\prime n}=w^{n}}}\left(  \cdot\right)  }_{\gamma
}+\underbrace{\sum_{\substack{u^{\prime n}\neq u^{n},\\v^{\prime n}\neq
v^{n},\\w^{\prime n}\neq w^{n}}}\left(  \cdot\right)  }_{\Delta}%
,\label{eq:sum-decomposition}%
\end{equation}
where we label the four different error terms as $\alpha$, $\beta$, $\gamma$,
and $\Delta$.

We now analyze each of these sums individually and introduce the expectation
$\mathbb{E}_{S^{n}}\ \mathbb{E}_{X_{1}^{n}|u^{n}S^{n},X_{2}^{n}|v^{n}S^{n}}$,
which corresponds to the expectations over the random choice of codebook.

\textbf{Error }$\alpha$\textbf{.} Consider the error term from the first sum:%
\begin{multline*}
4\sum_{u^{n},v^{n},w^{n}}p\left(  u^{n},v^{n},w^{n}\right)  \times\\
\sum_{u^{\prime n}\neq u^{n}}\mathbb{E}_{S^{n}}\ \mathbb{E}_{X_{1}^{n}%
|u^{n}S^{n},X_{2}^{n}|v^{n}S^{n}}\left\{  \text{Tr}\left\{  \Gamma_{u^{\prime
n},v^{n},w^{n}}\ \rho_{X_{1}^{n}\left(  u^{n},S^{n}\right)  ,X_{2}^{n}\left(
v^{n},S^{n}\right)  }^{\prime}\right\}  \right\}  \times\\
\mathcal{I}\left(  \left(  u^{n},v^{n},w^{n}\right)  \in\mathcal{T}%
^{n}\right)  \mathcal{I}\left(  \left(  u^{\prime n},v^{n},w^{n}\right)
\in\mathcal{T}^{n}\right)  .
\end{multline*}
Let us focus on bounding the middle part of the above expression:%
\begin{align*}
&\!\!\!\!\!\!\!  \sum_{u^{\prime n}\neq u^{n}}\mathbb{E}_{S^{n}}\ \mathbb{E}_{X_{1}%
^{n}|u^{n}S^{n},X_{2}^{n}|v^{n}S^{n}}\left\{  \text{Tr}\left\{  \Gamma
_{u^{\prime n},v^{n},w^{n}}\ \rho_{X_{1}^{n}\left(  u^{n},S^{n}\right)
,X_{2}^{n}\left(  v^{n},S^{n}\right)  }^{\prime}\right\}  \right\}  \\
&  =\sum_{u^{\prime n}\neq u^{n}}\mathbb{E}_{S^{n}}\ \mathbb{E}_{X_{1}%
^{n}|u^{n}S^{n},X_{2}^{n}|v^{n}S^{n}}\left\{  \text{Tr}\left\{  \Gamma
_{u^{\prime n},v^{n},w^{n}}\ \Pi_{X_{2}^{n}\left(  v^{n},S^{n}\right)  }%
\ \rho_{X_{1}^{n}\left(  u^{n},S^{n}\right)  ,X_{2}^{n}\left(  v^{n}%
,S^{n}\right)  }\ \Pi_{X_{2}^{n}\left(  v^{n},S^{n}\right)  }\right\}
\right\}  \\
&  =\sum_{u^{\prime n}\neq u^{n}}\mathbb{E}_{S^{n}}\ \mathbb{E}_{X_{2}%
^{n}|v^{n}S^{n}}\left\{  \text{Tr}\left\{  \Gamma_{u^{\prime n},v^{n},w^{n}%
}\ \Pi_{X_{2}^{n}\left(  v^{n},S^{n}\right)  }\ \mathbb{E}_{X_{1}^{n}%
|u^{n}S^{n}}\left\{  \rho_{X_{1}^{n}\left(  u^{n},S^{n}\right)  ,X_{2}%
^{n}\left(  v^{n},S^{n}\right)  }\right\}  \ \Pi_{X_{2}^{n}\left(  v^{n}%
,S^{n}\right)  }\right\}  \right\}  \\
&  =\sum_{u^{\prime n}\neq u^{n}}\mathbb{E}_{S^{n}}\ \mathbb{E}_{X_{2}%
^{n}|v^{n}S^{n}}\left\{  \text{Tr}\left\{  \Gamma_{u^{\prime n},v^{n},w^{n}%
}\ \Pi_{X_{2}^{n}\left(  v^{n},S^{n}\right)  }\ \rho_{X_{2}^{n}\left(
v^{n},S^{n}\right)  }^{\left(  u^{n}\right)  }\ \Pi_{X_{2}^{n}\left(
v^{n},S^{n}\right)  }\right\}  \right\}
\end{align*}
The first equality follows from the definition of $\rho_{X_{1}^{n}\left(
u^{n},S^{n}\right)  ,X_{2}^{n}\left(  v^{n},S^{n}\right)  }^{\prime}$. The
second equality follows because $u^{\prime n}\neq u^{n}$ and from the way that
the code was chosen randomly. The third equality is from the definition in
(\ref{eq:u-state}):%
\[
\mathbb{E}_{X_{1}^{n}|u^{n}S^{n}}\left\{  \rho_{X_{1}^{n}\left(  u^{n}%
,S^{n}\right)  ,X_{2}^{n}\left(  v^{n},S^{n}\right)  }\right\}  =\rho
_{X_{2}^{n}\left(  v^{n},S^{n}\right)  }^{\left(  u^{n}\right)  }.
\]
At this point, we substitute back into our earlier expression and continue
upper bounding:%
\begin{align*}
&  \sum_{u^{n},v^{n},w^{n}}p\left(  u^{n},v^{n},w^{n}\right)  \sum_{u^{\prime
n}\neq u^{n}}\mathbb{E}_{S^{n}}\ \mathbb{E}_{X_{2}^{n}|v^{n}S^{n}}\left\{
\text{Tr}\left\{  \Gamma_{u^{\prime n},v^{n},w^{n}}\ \Pi_{X_{2}^{n}\left(
v^{n},S^{n}\right)  }\ \rho_{X_{2}^{n}\left(  v^{n},S^{n}\right)  }^{\left(
u^{n}\right)  }\ \Pi_{X_{2}^{n}\left(  v^{n},S^{n}\right)  }\right\}
\right\}  \times\\
&  \ \ \ \ \ \ \ \ \ \ \ \ \ \ \ \mathcal{I}\left(  \left(  u^{n},v^{n}%
,w^{n}\right)  \in\mathcal{T}^{n}\right)  \mathcal{I}\left(  \left(  u^{\prime
n},v^{n},w^{n}\right)  \in\mathcal{T}^{n}\right)
\end{align*}
Consider the following inequality:%
\begin{align*}
\sum_{u^{n},v^{n},w^{n}}p\left(  u^{n},v^{n},w^{n}\right)   &  =\sum
_{u^{n},v^{n},w^{n}}p\left(  u^{n}|v^{n}\right)  p\left(  v^{n}\right)
\delta_{w^{n},f^{n}\left(  u^{n}\right)  }\delta_{w^{n},g^{n}\left(
v^{n}\right)  }\\
&  \leq\sum_{u^{n},v^{n},w^{n}}p\left(  u^{n}|v^{n}\right)  p\left(
v^{n}\right)  \delta_{w^{n},g^{n}\left(  v^{n}\right)  }%
\end{align*}
Substituting and distributing gives us the upper bound:%
\begin{multline*}
\leq\sum_{v^{n},w^{n}}p\left(  v^{n}\right)  \delta_{w^{n},g^{n}\left(
v^{n}\right)  }\times\\
\sum_{u^{\prime n}\neq u^{n}}\mathbb{E}_{S^{n}}\ \mathbb{E}_{X_{2}^{n}%
|v^{n}S^{n}}\left\{  \text{Tr}\left\{  \Gamma_{u^{\prime n},v^{n},w^{n}}%
\ \Pi_{X_{2}^{n}\left(  v^{n},S^{n}\right)  }\ \left(  \sum_{u^{n}}p\left(
u^{n}|v^{n}\right)  \rho_{X_{2}^{n}\left(  v^{n},S^{n}\right)  }^{\left(
u^{n}\right)  }\right)  \ \Pi_{X_{2}^{n}\left(  v^{n},S^{n}\right)  }\right\}
\right\}  \times\\
\mathcal{I}\left(  \left(  u^{n},v^{n},w^{n}\right)  \in\mathcal{T}%
^{n}\right)  \mathcal{I}\left(  \left(  u^{\prime n},v^{n},w^{n}\right)
\in\mathcal{T}^{n}\right)  .
\end{multline*}%
\begin{multline*}
\leq\sum_{v^{n},w^{n}}p\left(  v^{n}\right)  \delta_{w^{n},g^{n}\left(
v^{n}\right)  }\times\\
\sum_{u^{\prime n}\neq u^{n}}\mathbb{E}_{S^{n}}\ \mathbb{E}_{X_{2}^{n}%
|v^{n}S^{n}}\left\{  \text{Tr}\left\{  \Gamma_{u^{\prime n},v^{n},w^{n}}%
\ \Pi_{X_{2}^{n}\left(  v^{n},S^{n}\right)  }\ \left(  \sum_{u^{n}}p\left(
u^{n}|v^{n}\right)  \rho_{X_{2}^{n}\left(  v^{n},S^{n}\right)  }^{\left(
u^{n}\right)  }\right)  \ \Pi_{X_{2}^{n}\left(  v^{n},S^{n}\right)  }\right\}
\right\}  \times\\
\mathcal{I}\left(  \left(  u^{\prime n},v^{n},w^{n}\right)  \in\mathcal{T}%
^{n}\right)  ,
\end{multline*}
where the second inequality follows from dropping the indicator function
$\mathcal{I}\left(  \left(  u^{n},v^{n},w^{n}\right)  \in\mathcal{T}%
^{n}\right)  $. Focusing on the expression in the middle (and omitting the
indicator function for the moment), we have%
\begin{align*}
&  \sum_{u^{\prime n}\neq u^{n}}\mathbb{E}_{S^{n}}\ \mathbb{E}_{X_{2}%
^{n}|v^{n}S^{n}}\left\{  \text{Tr}\left\{  \Gamma_{u^{\prime n},v^{n},w^{n}%
}\ \Pi_{X_{2}^{n}\left(  v^{n},S^{n}\right)  }\ \left(  \sum_{u^{n}}p\left(
u^{n}|v^{n}\right)  \rho_{X_{2}^{n}\left(  v^{n},S^{n}\right)  }^{\left(
u^{n}\right)  }\right)  \ \Pi_{X_{2}^{n}\left(  v^{n},S^{n}\right)  }\right\}
\right\}  \\
&  =\sum_{u^{\prime n}\neq u^{n}}\mathbb{E}_{S^{n}}\ \mathbb{E}_{X_{2}%
^{n}|v^{n}S^{n}}\left\{  \text{Tr}\left\{  \Gamma_{u^{\prime n},v^{n},w^{n}%
}\ \Pi_{X_{2}^{n}\left(  v^{n},S^{n}\right)  }\ \sigma_{X_{2}^{n}\left(
v^{n},S^{n}\right)  }\ \Pi_{X_{2}^{n}\left(  v^{n},S^{n}\right)  }\right\}
\right\}  \\
&  \leq2^{-n\left[  H\left(  B|X_{2}VS\right)  -\delta\right]  }%
\sum_{u^{\prime n}\neq u^{n}}\mathbb{E}_{S^{n}}\ \mathbb{E}_{X_{2}^{n}%
|v^{n}S^{n}}\left\{  \text{Tr}\left\{  \Gamma_{u^{\prime n},v^{n},w^{n}}%
\ \Pi_{X_{2}^{n}\left(  v^{n},S^{n}\right)  }\right\}  \right\}  \\
&  =2^{-n\left[  H\left(  B|X_{2}VS\right)  -\delta\right]  }\sum_{u^{\prime
n}\neq u^{n}}\mathbb{E}_{S^{n}}\ \mathbb{E}_{X_{2}^{n}|v^{n}S^{n}}\left\{
\text{Tr}\left\{
\begin{array}
[c]{c}%
\Pi\ \Pi_{S^{n}\left(  w^{n}\right)  }\ \Pi_{X_{1}^{n}\left(  u^{\prime
n},S^{n}\right)  }\ \Pi_{X_{1}^{n}\left(  u^{\prime n},S^{n}\right)
,X_{2}^{n}\left(  v^{n},S^{n}\right)  }\ \times\\
\Pi_{X_{1}^{n}\left(  u^{\prime n},S^{n}\right)  }\ \Pi_{S^{n}\left(
w^{n}\right)  }\ \Pi\ \Pi_{X_{2}^{n}\left(  v^{n},S^{n}\right)  }%
\end{array}
\right\}  \right\}  \\
&  \leq2^{-n\left[  H\left(  B|X_{2}VS\right)  -\delta\right]  }%
\sum_{u^{\prime n}\neq u^{n}}\mathbb{E}_{S^{n}}\ \mathbb{E}_{X_{2}^{n}%
|v^{n}S^{n}}\left\{  \text{Tr}\left\{  \Pi_{X_{1}^{n}\left(  u^{\prime
n},S^{n}\right)  ,X_{2}^{n}\left(  v^{n},S^{n}\right)  }\right\}  \right\}  \\
&  \leq2^{-n\left[  H\left(  B|X_{2}VS\right)  -\delta\right]  }2^{n\left[
H\left(  B|X_{1}X_{2}\right)  +\delta\right]  }\sum_{u^{\prime n}\neq u^{n}}1
\end{align*}
The first equality follows from the definition in (\ref{eq:sigma-state}):%
\[
\sigma_{X_{2}^{n}\left(  v^{n},S^{n}\right)  }=\sum_{u^{n}}p\left(
u^{n}|v^{n}\right)  \rho_{X_{2}^{n}\left(  v^{n},S^{n}\right)  }^{\left(
u^{n}\right)  }.
\]
The first inequality follows from the property (\ref{eq:property-3-typical})
of conditionally typical projectors, applied to the state $\sigma_{X_{2}%
^{n}\left(  v^{n},S^{n}\right)  }$ and its corresponding typical projector
$\Pi_{X_{2}^{n}\left(  v^{n},S^{n}\right)  }$. The second equality follows
from the definition (\ref{eq:gamma-def}) of $\Gamma_{u^{\prime n},v^{n},w^{n}%
}$. The second inequality is from cyclicity of trace and the fact that $\Pi$,
$\Pi_{S^{n}\left(  w^{n}\right)  }$, $\Pi_{X_{1}^{n}\left(  u^{\prime n}%
,S^{n}\right)  }\leq I$. The final inequality is from the property
(\ref{eq:property-2-typical}) of typical projectors (for the projector
$\Pi_{X_{1}^{n}\left(  u^{\prime n},S^{n}\right)  ,X_{2}^{n}\left(
v^{n},S^{n}\right)  }$). Substituting back in, we obtain the upper bound%
\[
\leq2^{-n\left[  H\left(  B|X_{2}VS\right)  -\delta\right]  }2^{n\left[
H\left(  B|X_{1}X_{2}\right)  +\delta\right]  }\sum_{v^{n},w^{n}}p\left(
v^{n}\right)  \delta_{w^{n},g^{n}\left(  v^{n}\right)  }\sum_{u^{\prime n}\neq
u^{n}}\mathcal{I}\left(  \left(  u^{\prime n},v^{n},w^{n}\right)
\in\mathcal{T}^{n}\right)  ,
\]
which we can further upper bound as%
\begin{align*}
&  \leq2^{-n\left[  H\left(  B|X_{2}VS\right)  -\delta\right]  }2^{n\left[
H\left(  B|X_{1}X_{2}\right)  +\delta\right]  }\sum_{v^{n},w^{n}}p\left(
v^{n}\right)  \delta_{w^{n},g^{n}\left(  v^{n}\right)  }\sum_{u^{\prime n}\neq
u^{n}}p\left(  u^{\prime n}|v^{n},w^{n}\right)  2^{n\left[  H\left(
U|VW\right)  +\delta^{\prime}\right]  }\\
&  \leq2^{-n\left[  H\left(  B|X_{2}VS\right)  -\delta\right]  }2^{n\left[
H\left(  B|X_{1}X_{2}\right)  +\delta\right]  }2^{n\left[  H\left(
U|VW\right)  +\delta^{\prime}\right]  }\\
&  =2^{-n\left[  I\left(  X_{1};B|X_{2}VS\right)  -H\left(  U|V\right)
-2\delta-\delta^{\prime}\right]  },
\end{align*}
where we exploited the fact that%
\[
\mathcal{I}\left(  \left(  u^{\prime n},v^{n},w^{n}\right)  \in\mathcal{T}%
^{n}\right)  \leq p\left(  u^{\prime n}|v^{n},w^{n}\right)  2^{n\left[
H\left(  U|VW\right)  +\delta^{\prime}\right]  }.
\]
(See Appendix~\ref{sec:indicator-function-trick}\ for a discussion of this
fact.) Substituting this back into our original expression, we have the bound:%
\[
\leq4\ 2^{-n\left[  I\left(  X_{1};B|X_{2}VS\right)  -H\left(  UW|V\right)
-2\delta-\delta^{\prime}\right]  }.
\]

\textbf{Error }$\beta$\textbf{.} We now handle the error from the second sum:%
\begin{multline*}
4\sum_{u^{n},v^{n},w^{n}}p\left(  u^{n},v^{n},w^{n}\right)  \times\\
\sum_{v^{\prime n}\neq v^{n}}\mathbb{E}_{S^{n}}\ \mathbb{E}_{X_{1}^{n}%
|u^{n}S^{n},X_{2}^{n}|v^{n}S^{n}}\left\{  \text{Tr}\left\{  \Gamma
_{u^{n},v^{\prime n},w^{n}}\ \rho_{X_{1}^{n}\left(  u^{n},S^{n}\right)
,X_{2}^{n}\left(  v^{n},S^{n}\right)  }^{\prime}\right\}  \right\}  \times\\
\mathcal{I}\left(  \left(  u^{n},v^{n},w^{n}\right)  \in\mathcal{T}%
^{n}\right)  \mathcal{I}\left(  \left(  u^{n},v^{\prime n},w^{n}\right)
\in\mathcal{T}^{n}\right)  .
\end{multline*}
Let us again focus on bounding the middle part:%
\[
\sum_{v^{\prime n}\neq v^{n}}\mathbb{E}_{S^{n}}\ \mathbb{E}_{X_{1}^{n}%
|u^{n}S^{n},X_{2}^{n}|v^{n}S^{n}}\left\{  \text{Tr}\left\{  \Gamma
_{u^{n},v^{\prime n},w^{n}}\ \rho_{X_{1}^{n}\left(  u^{n},S^{n}\right)
,X_{2}^{n}\left(  v^{n},S^{n}\right)  }^{\prime}\right\}  \right\}
\]
Expanding $\mathbb{E}_{X_{2}^{n}|v^{\prime n}S^{n}}\left\{  \Gamma
_{u^{n},v^{\prime n},w^{n}}\right\}  $ as%
\begin{align*}
&  \mathbb{E}_{X_{2}^{n}|v^{\prime n}S^{n}}\left\{  \Gamma_{u^{n},v^{\prime
n},w^{n}}\right\}  \\
&  =\mathbb{E}_{X_{2}^{n}|v^{\prime n}S^{n}}\left\{  \Pi\ \Pi_{S^{n}\left(
w^{n}\right)  }\ \Pi_{X_{1}^{n}\left(  u^{n},S^{n}\right)  }\ \Pi_{X_{1}%
^{n}\left(  u^{n},S^{n}\right)  ,X_{2}^{n}\left(  v^{\prime n},S^{n}\right)
}\ \Pi_{X_{1}^{n}\left(  u^{n},S^{n}\right)  }\ \Pi_{S^{n}\left(
w^{n}\right)  }\ \Pi\right\}  \\
&  =\Pi\ \Pi_{S^{n}\left(  w^{n}\right)  }\ \Pi_{X_{1}^{n}\left(  u^{n}%
,S^{n}\right)  }\ \mathbb{E}_{X_{2}^{n}|v^{\prime n}S^{n}}\left\{  \Pi
_{X_{1}^{n}\left(  u^{n},S^{n}\right)  ,X_{2}^{n}\left(  v^{\prime n}%
,S^{n}\right)  }\right\}  \ \Pi_{X_{1}^{n}\left(  u^{n},S^{n}\right)  }%
\ \Pi_{S^{n}\left(  w^{n}\right)  }\ \Pi\\
&  \leq2^{n\left[  H\left(  B|X_{1}X_{2}\right)  +\delta\right]  }\ \Pi
\ \Pi_{S^{n}\left(  w^{n}\right)  }\ \Pi_{X_{1}^{n}\left(  u^{n},S^{n}\right)
}\ \mathbb{E}_{X_{2}^{n}|v^{\prime n}S^{n}}\left\{  \rho_{X_{1}^{n}\left(
u^{n},S^{n}\right)  ,X_{2}^{n}\left(  v^{\prime n},S^{n}\right)  }\right\}
\ \Pi_{X_{1}^{n}\left(  u^{n},S^{n}\right)  }\ \Pi_{S^{n}\left(  w^{n}\right)
}\ \Pi\\
&  =2^{n\left[  H\left(  B|X_{1}X_{2}\right)  +\delta\right]  }\ \Pi
\ \Pi_{S^{n}\left(  w^{n}\right)  }\ \Pi_{X_{1}^{n}\left(  u^{n},S^{n}\right)
}\ \rho_{X_{1}^{n}\left(  u^{n},S^{n}\right)  }^{\left(  v^{\prime n}\right)
}\ \Pi_{X_{1}^{n}\left(  u^{n},S^{n}\right)  }\ \Pi_{S^{n}\left(
w^{n}\right)  }\ \Pi.
\end{align*}
The first equality is a definition. The second equality follows because
$v^{\prime n}\neq v^{n}$ for terms in the sum and from the way that the code
was chosen. The first inequality is the \textquotedblleft projector
trick\textquotedblright\ discussed in Appendix~\ref{sec:typ-review}. The final
equality is from the definition in (\ref{eq:v-state}). Substituting back in
gives the upper bound%
\begin{multline*}
\leq2^{n\left[  H\left(  B|X_{1}X_{2}\right)  +\delta\right]  }\sum_{v^{\prime
n}\neq v^{n}}\mathbb{E}_{S^{n}}\ \mathbb{E}_{\substack{X_{1}^{n}|u^{n}%
S^{n},\\X_{2}^{n}|v^{n}S^{n}}}\left\{  \text{Tr}\left\{
\begin{array}
[c]{c}%
\Pi\ \Pi_{S^{n}\left(  w^{n}\right)  }\ \Pi_{X_{1}^{n}\left(  u^{n}%
,S^{n}\right)  }\ \rho_{X_{1}^{n}\left(  u^{n},S^{n}\right)  }^{\left(
v^{\prime n}\right)  }\ \Pi_{X_{1}^{n}\left(  u^{n},S^{n}\right)  }\times\\
\Pi_{S^{n}\left(  w^{n}\right)  }\ \Pi\ \rho_{X_{1}^{n}\left(  u^{n}%
,S^{n}\right)  ,X_{2}^{n}\left(  v^{n},S^{n}\right)  }^{\prime}%
\end{array}
\right\}  \right\}  \times\\
\mathcal{I}\left(  \left(  u^{n},v^{n},w^{n}\right)  \in\mathcal{T}%
^{n}\right)  \mathcal{I}\left(  \left(  u^{n},v^{\prime n},w^{n}\right)
\in\mathcal{T}^{n}\right)
\end{multline*}
We now note the upper bounds%
\begin{align*}
\mathcal{I}\left(  \left(  u^{n},v^{n},w^{n}\right)  \in\mathcal{T}%
^{n}\right)  \mathcal{I}\left(  \left(  u^{n},v^{\prime n},w^{n}\right)
\in\mathcal{T}^{n}\right)   &  \leq\mathcal{I}\left(  \left(  u^{n},v^{\prime
n}\right)  \in\mathcal{T}^{n}\right)  \\
&  \leq p\left(  v^{\prime n}|u^{n}\right)  2^{n\left[  H\left(  V|U\right)
+\delta^{\prime}\right]  }%
\end{align*}
which upon substitution and distribution give the upper bound%
\begin{align*}
&  \leq2^{n\left[  H\left(  B|X_{1}X_{2}\right)  +H\left(  V|U\right)
+\delta+\delta^{\prime}\right]  }\mathbb{E}_{S^{n}}\ \mathbb{E}%
_{\substack{X_{1}^{n}|u^{n}S^{n},\\X_{2}^{n}|v^{n}S^{n}}}\left\{
\text{Tr}\left\{
\begin{array}
[c]{c}%
\Pi\ \Pi_{S^{n}\left(  w^{n}\right)  }\ \times\\
\Pi_{X_{1}^{n}\left(  u^{n},S^{n}\right)  }\ \left(  \sum_{v^{\prime n}%
}p\left(  v^{\prime n}|u^{n}\right)  \rho_{X_{1}^{n}\left(  u^{n}%
,S^{n}\right)  }^{\left(  v^{\prime n}\right)  }\right)  \ \Pi_{X_{1}%
^{n}\left(  u^{n},S^{n}\right)  }\times\\
\ \Pi_{S^{n}\left(  w^{n}\right)  }\ \Pi\ \rho_{X_{1}^{n}\left(  u^{n}%
,S^{n}\right)  ,X_{2}^{n}\left(  v^{n},S^{n}\right)  }^{\prime}%
\end{array}
\right\}  \right\}  \\
&  \leq2^{n\left[  H\left(  B|X_{1}X_{2}\right)  +\delta\right]  }2^{n\left[
H\left(  V|U\right)  +\delta^{\prime}\right]  }2^{-n\left[  H\left(
B|X_{1}US\right)  -\delta\right]  }\mathbb{E}_{S^{n}}\ \mathbb{E}%
_{\substack{X_{1}^{n}|u^{n}S^{n},\\X_{2}^{n}|v^{n}S^{n}}}\left\{
\text{Tr}\left\{
\begin{array}
[c]{c}%
\Pi\ \Pi_{S^{n}\left(  w^{n}\right)  }\ \Pi_{X_{1}^{n}\left(  u^{n}%
,S^{n}\right)  }\times\\
\ \Pi_{S^{n}\left(  w^{n}\right)  }\ \Pi\ \rho_{X_{1}^{n}\left(  u^{n}%
,S^{n}\right)  ,X_{2}^{n}\left(  v^{n},S^{n}\right)  }^{\prime}%
\end{array}
\right\}  \right\}  \\
&  \leq2^{n\left[  H\left(  B|X_{1}X_{2}\right)  +\delta\right]  }2^{n\left[
H\left(  V|U\right)  +\delta^{\prime}\right]  }2^{-n\left[  H\left(
B|X_{1}US\right)  -\delta\right]  }\\
&  =2^{-n\left[  I\left(  X_{2};B|X_{1}US\right)  -H\left(  V|U\right)
-\delta^{\prime}-2\delta\right]  }%
\end{align*}
Substituting back in then gives the bound:%
\begin{align*}
&  4\ 2^{-n\left[  I\left(  X_{2};B|X_{1}US\right)  -H\left(  V|U\right)
-\delta^{\prime}-2\delta\right]  }\sum_{u^{n},v^{n},w^{n}}p\left(  u^{n}%
,v^{n},w^{n}\right)  \\
&  \leq4\ 2^{-n\left[  I\left(  X_{2};B|X_{1}US\right)  -H\left(  V|U\right)
-\delta^{\prime}-2\delta\right]  }.
\end{align*}

\textbf{Error }$\gamma$\textbf{.} We now handle the error from the third sum
(this error corresponds to the common part $w^{n}$ being decoded correctly,
but $u^{n}$ and $v^{n}$ both being decoded incorrectly):%
\begin{multline*}
4\sum_{u^{n},v^{n},w^{n}}p\left(  u^{n},v^{n},w^{n}\right)  \times\\
\sum_{u^{\prime n}\neq u^{n},v^{\prime n}\neq v^{n}}\mathbb{E}_{S^{n}%
}\ \mathbb{E}_{X_{1}^{n}|u^{n}S^{n},X_{2}^{n}|v^{n}S^{n}}\left\{
\text{Tr}\left\{  \Gamma_{u^{\prime n},v^{\prime n},w^{n}}\ \rho_{X_{1}%
^{n}\left(  u^{n},S^{n}\right)  ,X_{2}^{n}\left(  v^{n},S^{n}\right)
}^{\prime}\right\}  \right\}  \times\\
\mathcal{I}\left(  \left(  u^{n},v^{n},w^{n}\right)  \in\mathcal{T}%
^{n}\right)  \mathcal{I}\left(  \left(  u^{\prime n},v^{\prime n}%
,w^{n}\right)  \in\mathcal{T}^{n}\right)  .
\end{multline*}
Again consider the middle part:%
\begin{align*}
&  \sum_{\substack{u^{\prime n}\neq u^{n},\\v^{\prime n}\neq v^{n}}%
}\mathbb{E}_{S^{n}}\ \mathbb{E}_{X_{1}^{n}|u^{n}S^{n},X_{2}^{n}|v^{n}S^{n}%
}\left\{  \text{Tr}\left\{  \Gamma_{u^{\prime n},v^{\prime n},w^{n}}%
\ \rho_{X_{1}^{n}\left(  u^{n},S^{n}\right)  ,X_{2}^{n}\left(  v^{n}%
,S^{n}\right)  }^{\prime}\right\}  \right\}  \\
&  =\sum_{\substack{u^{\prime n}\neq u^{n},\\v^{\prime n}\neq v^{n}%
}}\mathbb{E}_{S^{n}}\ \mathbb{E}_{X_{1}^{n}|u^{n}S^{n},X_{2}^{n}|v^{n}S^{n}%
}\left\{  \text{Tr}\left\{
\begin{array}
[c]{c}%
\Pi\ \Pi_{S^{n}\left(  w^{n}\right)  }\ \Pi_{X_{1}^{n}\left(  u^{\prime
n},S^{n}\right)  }\ \Pi_{X_{1}^{n}\left(  u^{\prime n},S^{n}\right)
,X_{2}^{n}\left(  v^{\prime n},S^{n}\right)  }\times\\
\Pi_{X_{1}^{n}\left(  u^{\prime n},S^{n}\right)  }\ \Pi_{S^{n}\left(
w^{n}\right)  }\ \Pi\ \rho_{X_{1}^{n}\left(  u^{n},S^{n}\right)  ,X_{2}%
^{n}\left(  v^{n},S^{n}\right)  }^{\prime}%
\end{array}
\right\}  \right\}  \\
&  \leq2^{n\left[  H\left(  B|X_{1}X_{2}\right)  +\delta\right]  }%
\sum_{\substack{u^{\prime n}\neq u^{n},\\v^{\prime n}\neq v^{n}}%
}\mathbb{E}_{S^{n}}\ \mathbb{E}_{X_{1}^{n}|u^{n}S^{n},X_{2}^{n}|v^{n}S^{n}%
}\left\{  \text{Tr}\left\{
\begin{array}
[c]{c}%
\Pi\ \Pi_{S^{n}\left(  w^{n}\right)  }\ \Pi_{X_{1}^{n}\left(  u^{\prime
n},S^{n}\right)  }\ \rho_{X_{1}^{n}\left(  u^{\prime n},S^{n}\right)
,X_{2}^{n}\left(  v^{\prime n},S^{n}\right)  }\times\\
\Pi_{X_{1}^{n}\left(  u^{\prime n},S^{n}\right)  }\ \Pi_{S^{n}\left(
w^{n}\right)  }\ \Pi\ \rho_{X_{1}^{n}\left(  u^{n},S^{n}\right)  ,X_{2}%
^{n}\left(  v^{n},S^{n}\right)  }^{\prime}%
\end{array}
\right\}  \right\}
\end{align*}
Here, we have again used the \textquotedblleft projector trick
inequality\textquotedblright:%
\[
\Pi_{X_{1}^{n}\left(  u^{\prime n},S^{n}\right)  ,X_{2}^{n}\left(  v^{\prime
n},S^{n}\right)  }\leq2^{n\left[  H\left(  B|X_{1}X_{2}\right)  +\delta
\right]  }\ \rho_{X_{1}^{n}\left(  u^{\prime n},S^{n}\right)  ,X_{2}%
^{n}\left(  v^{\prime n},S^{n}\right)  }.
\]
Now we introduce the expectations over the random
choice of codewords.
Since $v^{\prime n} \neq v^{n}$, this means that there are two
independent random codewords involved, namely 
$X_{2}^{n}\left(  v^{n},S^{n}\left(  w^{n}\right)  \right)$ 
and $X_{2}^{n}\left(  v^{\prime n},S^{n}\left(  w^{n}\right)  \right)$. 
We take the expectation $\mathbb{E}_{X_{2}^{n}|v^{\prime n}S^{n}}$
inside the trace and obtain
\[
2^{n\left[  H\left(  B|X_{1}X_{2}\right)  +\delta\right]  }\sum
_{\substack{u^{\prime n}\neq u^{n},\\v^{\prime n}\neq v^{n}}}\mathbb{E}%
_{S^{n}}\ \mathbb{E}_{X_{1}^{n}|u^{n}S^{n},X_{2}^{n}|v^{n}S^{n}}\left\{
\text{Tr}\left\{
\begin{array}
[c]{c}%
\Pi\ \Pi_{S^{n}\left(  w^{n}\right)  }\ \Pi_{X_{1}^{n}\left(  u^{\prime
n},S^{n}\right)  }\ \rho_{X_{1}^{n}\left(  u^{\prime n},S^{n}\right)
}^{\left(  v^{\prime n}\right)  }\ \Pi_{X_{1}^{n}\left(  u^{\prime n}%
,S^{n}\right)  }\ \times\\
\Pi_{S^{n}\left(  w^{n}\right)  }\ \Pi\ \rho_{X_{1}^{n}\left(  u^{n}%
,S^{n}\right)  ,X_{2}^{n}\left(  v^{n},S^{n}\right)  }^{\prime}%
\end{array}
\right\}  \right\}
\]
We will also exploit the following \textquotedblleft indicator function
trick\textquotedblright\ in order to bound the full error event probability:%
\begin{align*}
\mathcal{I}\left(  \left(  u^{n},v^{n},w^{n}\right)  \in\mathcal{T}%
^{n}\right)  \mathcal{I}\left(  \left(  u^{\prime n},v^{\prime n}%
,w^{n}\right)  \in\mathcal{T}^{n}\right)   &  \leq\mathcal{I}\left(  \left(
u^{\prime n},v^{\prime n},w^{n}\right)  \in\mathcal{T}^{n}\right)  \\
&  \leq2^{n\left[  H\left(  UV|W\right)  +\delta^{\prime}\right]  }p\left(
u^{\prime n},v^{\prime n}|w^{n}\right)  \\
&  =2^{n\left[  H\left(  UV|W\right)  +\delta^{\prime}\right]  }p\left(
u^{\prime n}|w^{n}\right)  p\left(  v^{\prime n}|u^{\prime n},w^{n}\right)  \\
&  =2^{n\left[  H\left(  UV|W\right)  +\delta^{\prime}\right]  }p\left(
u^{\prime n}|w^{n}\right)  p\left(  v^{\prime n}|u^{\prime n}\right)  ,
\end{align*}
where the last equality holds because $p\left(  v^{\prime n}|u^{\prime
n},w^{n}\right)  =p\left(  v^{\prime n}|u^{\prime n}\right)  $ whenever
$w^{n}=f\left(  u^{n}\right)  =g\left(  v^{n}\right)  =f\left(  u^{\prime
n}\right)  =g\left(  v^{\prime n}\right)  $ (which we know is true for terms
in this third sum---otherwise, the delta function from $p\left(  u^{n}%
,v^{n},w^{n}\right)  $ ensures that the whole expression vanishes). Combining
the results of the projector trick inequality and the indicator function
trick, we then get the upper bound%
\begin{align*}
&  \leq2^{n\left[  H\left(  B|X_{1}X_{2}\right)  +\delta\right]  }2^{n\left[
H\left(  UV|W\right)  +\delta^{\prime}\right]  }\sum_{\substack{u^{\prime
n}\neq u^{n},\\v^{\prime n}\neq v^{n}}}p\left(  u^{\prime n}|w^{n}\right)
p\left(  v^{\prime n}|u^{\prime n}\right)  \times\\
&  \ \ \ \ \ \ \ \mathbb{E}_{S^{n}}\ \mathbb{E}_{X_{1}^{n}|u^{n}S^{n}%
,X_{2}^{n}|v^{n}S^{n}}\left\{  \text{Tr}\left\{
\begin{array}
[c]{c}%
\Pi\ \Pi_{S^{n}\left(  w^{n}\right)  }\ \Pi_{X_{1}^{n}\left(  u^{\prime
n},S^{n}\right)  }\ \rho_{X_{1}^{n}\left(  u^{\prime n},S^{n}\right)
}^{\left(  v^{\prime n}\right)  }\ \Pi_{X_{1}^{n}\left(  u^{\prime n}%
,S^{n}\right)  }\ \times\\
\Pi_{S^{n}\left(  w^{n}\right)  }\ \Pi\ \rho_{X_{1}^{n}\left(  u^{n}%
,S^{n}\right)  ,X_{2}^{n}\left(  v^{n},S^{n}\right)  }^{\prime}%
\end{array}
\right\}  \right\}  \\
&  \leq2^{n\left[  H\left(  B|X_{1}X_{2}\right)  +\delta\right]  }2^{n\left[
H\left(  UV|W\right)  +\delta^{\prime}\right]  }\sum_{u^{\prime n},v^{\prime
n}}p\left(  u^{\prime n}|w^{n}\right)  \times\\
&  \ \ \ \ \ \ \mathbb{E}_{S^{n}}\ \mathbb{E}_{X_{1}^{n}|u^{n}S^{n},X_{2}%
^{n}|v^{n}S^{n}}\left\{  \text{Tr}\left\{
\begin{array}
[c]{c}%
\Pi\ \Pi_{S^{n}\left(  w^{n}\right)  }\ \Pi_{X_{1}^{n}\left(  u^{\prime
n},S^{n}\right)  }\ \left(  \sum_{v^{\prime n}}p\left(  v^{\prime n}|u^{\prime
n}\right)  \rho_{X_{1}^{n}\left(  u^{\prime n},S^{n}\right)  }^{\left(
v^{\prime n}\right)  }\right)  \times\\
\Pi_{X_{1}^{n}\left(  u^{\prime n},S^{n}\right)  }\ \Pi_{S^{n}\left(
w^{n}\right)  }\ \Pi\ \rho_{X_{1}^{n}\left(  u^{n},S^{n}\right)  ,X_{2}%
^{n}\left(  v^{n},S^{n}\right)  }^{\prime}%
\end{array}
\right\}  \right\}
\end{align*}
Continuing, we have%
\begin{align*}
&  \leq2^{n\left[  H\left(  B|X_{1}X_{2}\right)  +\delta\right]  }2^{n\left[
H\left(  UV|W\right)  +\delta^{\prime}\right]  }\sum_{u^{\prime n}}p\left(
u^{\prime n}|w^{n}\right)  \times\\
&  \ \ \ \ \mathbb{E}_{S^{n}}\ \mathbb{E}_{X_{1}^{n}|u^{n}S^{n},X_{2}%
^{n}|v^{n}S^{n}}\left\{  \text{Tr}\left\{  \Pi\ \Pi_{S^{n}\left(
w^{n}\right)  }\ \left(  \sum_{v^{\prime n}}p\left(  v^{\prime n}|u^{\prime
n}\right)  \rho_{X_{1}^{n}\left(  u^{\prime n},S^{n}\right)  }^{\left(
v^{\prime n}\right)  }\right)  \ \Pi_{S^{n}\left(  w^{n}\right)  }\ \Pi
\ \rho_{X_{1}^{n}\left(  u^{n},S^{n}\right)  ,X_{2}^{n}\left(  v^{n}%
,S^{n}\right)  }^{\prime}\right\}  \right\}  \\
&  =2^{n\left[  H\left(  B|X_{1}X_{2}\right)  +\delta\right]  }2^{n\left[
H\left(  UV|W\right)  +\delta^{\prime}\right]  }\times\\
&  \ \ \ \ \mathbb{E}_{S^{n}}\ \mathbb{E}_{\substack{X_{1}^{n}|u^{n}%
S^{n},\\X_{2}^{n}|v^{n}S^{n}}}\left\{  \text{Tr}\left\{
\begin{array}
[c]{c}%
\Pi\ \Pi_{S^{n}\left(  w^{n}\right)  }\ \left(  \sum_{u^{\prime n},v^{\prime
n}}p\left(  u^{\prime n}|w^{n}\right)  \mathbb{E}_{X_{1}^{n}|u^{\prime n}%
S^{n}}\left\{  p\left(  v^{\prime n}|u^{\prime n}\right)  \rho_{X_{1}%
^{n}\left(  u^{\prime n},S^{n}\right)  }^{\left(  v^{\prime n}\right)
}\right\}  \right)  \ \Pi_{S^{n}\left(  w^{n}\right)  }\times\\
\Pi\ \rho_{X_{1}^{n}\left(  u^{n},S^{n}\right)  ,X_{2}^{n}\left(  v^{n}%
,S^{n}\right)  }^{\prime}%
\end{array}
\right\}  \right\}  \\
&  =2^{n\left[  H\left(  B|X_{1}X_{2}\right)  +\delta\right]  }2^{n\left[
H\left(  UV|W\right)  +\delta^{\prime}\right]  }\ \mathbb{E}_{S^{n}%
}\ \mathbb{E}_{\substack{X_{1}^{n}|u^{n}S^{n},\\X_{2}^{n}|v^{n}S^{n}}}\left\{
\text{Tr}\left\{
\begin{array}
[c]{c}%
\Pi\ \Pi_{S^{n}\left(  w^{n}\right)  }\ \tau_{S^{n}\left(  w^{n}\right)
}\ \Pi_{S^{n}\left(  w^{n}\right)  }\times\\
\Pi\ \rho_{X_{1}^{n}\left(  u^{n},S^{n}\right)  ,X_{2}^{n}\left(  v^{n}%
,S^{n}\right)  }^{\prime}%
\end{array}
\right\}  \right\}  \\
&  \leq2^{n\left[  H\left(  B|X_{1}X_{2}\right)  +\delta\right]  }2^{n\left[
H\left(  UV|W\right)  +\delta^{\prime}\right]  }2^{-n\left[  H\left(
B|WS\right)  -\delta\right]  }\mathbb{E}_{S^{n}}\ \mathbb{E}_{X_{1}^{n}%
|u^{n}S^{n},X_{2}^{n}|v^{n}S^{n}}\left\{  \text{Tr}\left\{  \Pi\ \Pi
_{S^{n}\left(  w^{n}\right)  }\ \Pi\ \rho_{X_{1}^{n}\left(  u^{n}%
,S^{n}\right)  ,X_{2}^{n}\left(  v^{n},S^{n}\right)  }^{\prime}\right\}
\right\}  \\
&  \leq2^{n\left[  H\left(  B|X_{1}X_{2}\right)  +\delta\right]  }2^{n\left[
H\left(  UV|W\right)  +\delta^{\prime}\right]  }2^{-n\left[  H\left(
B|WS\right)  -\delta\right]  }\\
&  =2^{-n\left[  I\left(  X_{1}X_{2};B|WS\right)  -H\left(  UV|W\right)
-2\delta-\delta^{\prime}\right]  }.
\end{align*}
The first inequality follows from the fact that%
\begin{align*}
& \Pi_{X_{1}^{n}\left(  u^{\prime n},S^{n}\right)  }\ \left(  \sum_{v^{\prime
n}}p\left(  v^{\prime n}|u^{\prime n}\right)  \rho_{X_{1}^{n}\left(  u^{\prime
n},S^{n}\right)  }^{\left(  v^{\prime n}\right)  }\right)  \ \Pi_{X_{1}%
^{n}\left(  u^{\prime n},S^{n}\right)  }\\
& =\left(  \sum_{v^{\prime n}}p\left(  v^{\prime n}|u^{\prime n}\right)
\rho_{X_{1}^{n}\left(  u^{\prime n},S^{n}\right)  }^{\left(  v^{\prime
n}\right)  }\right)  ^{1/2}\ \Pi_{X_{1}^{n}\left(  u^{\prime n},S^{n}\right)
}\ \left(  \sum_{v^{\prime n}}p\left(  v^{\prime n}|u^{\prime n}\right)
\rho_{X_{1}^{n}\left(  u^{\prime n},S^{n}\right)  }^{\left(  v^{\prime
n}\right)  }\right)  ^{1/2}\\
& \leq\sum_{v^{\prime n}}p\left(  v^{\prime n}|u^{\prime n}\right)
\rho_{X_{1}^{n}\left(  u^{\prime n},S^{n}\right)  }^{\left(  v^{\prime
n}\right)  }.
\end{align*}
The first equality follows from bringing the sum $\sum_{u^{\prime n}}p\left(
u^{\prime n}|w^{n}\right)  $ and expectation $\mathbb{E}_{X_{1}^{n}|u^{\prime
n}S^{n}}$ inside the trace. The second equality follows from the definition in
(\ref{eq:tau-state}):%
\[
\tau_{S^{n}\left(  w^{n}\right)  }\equiv\sum_{u^{\prime n},v^{\prime n}%
}p\left(  u^{\prime n}|w^{n}\right)  \mathbb{E}_{X_{1}^{n}|u^{\prime n}S^{n}%
}\left\{  p\left(  v^{\prime n}|u^{\prime n}\right)  \rho_{X_{1}^{n}\left(
u^{\prime n},S^{n}\right)  }^{\left(  v^{\prime n}\right)  }\right\}  .
\]
The second inequality follows from the property (\ref{eq:property-3-typical})
of conditionally typical projectors:%
\[
\Pi_{S^{n}\left(  w^{n}\right)  }\ \tau_{S^{n}\left(  w^{n}\right)  }%
\ \Pi_{S^{n}\left(  w^{n}\right)  }\leq2^{-n\left[  H\left(  B|WS\right)
-\delta\right]  }\ \Pi_{S^{n}\left(  w^{n}\right)  }.
\]
The final inequality follows from the bound:%
\[
\text{Tr}\left\{  \Pi\ \Pi_{S^{n}\left(  w^{n}\right)  }\ \Pi\ \rho_{X_{1}%
^{n}\left(  u^{n},S^{n}\right)  ,X_{2}^{n}\left(  v^{n},S^{n}\right)
}^{\prime}\right\}  \leq1.
\]

\textbf{Error }$\Delta$\textbf{.} Finally, we upper bound the error from the
last sum, which corresponds to all components of the source being erroneously
decoded:%
\begin{multline*}
4\sum_{u^{n},v^{n},w^{n}}p\left(  u^{n},v^{n},w^{n}\right)  \times\\
\sum_{u^{\prime n}\neq u^{n},v^{\prime n}\neq v^{n},w^{\prime n}\neq w^{n}%
}\mathbb{E}_{S^{n}}\ \mathbb{E}_{X_{1}^{n}|u^{n}S^{n},X_{2}^{n}|v^{n}S^{n}%
}\left\{  \text{Tr}\left\{  \Gamma_{u^{\prime n},v^{\prime n},w^{\prime n}%
}\ \rho_{X_{1}^{n}\left(  u^{n},S^{n}\right)  ,X_{2}^{n}\left(  v^{n}%
,S^{n}\right)  }^{\prime}\right\}  \right\}  \times\\
\mathcal{I}\left(  \left(  u^{n},v^{n},w^{n}\right)  \in\mathcal{T}%
^{n}\right)  \mathcal{I}\left(  \left(  u^{\prime n},v^{\prime n},w^{\prime
n}\right)  \in\mathcal{T}^{n}\right)  .
\end{multline*}%
\begin{multline*}
\leq4\sum_{u^{\prime n},v^{\prime n},w^{\prime n}}\sum_{u^{n},v^{n},w^{n}%
}p\left(  v^{n}\right)  p\left(  u^{n},w^{n}|v^{n}\right)  \times\\
\text{Tr}\left\{  \Gamma_{u^{\prime n},v^{\prime n},w^{\prime n}}%
\ \mathbb{E}_{S^{n}}\ \mathbb{E}_{X_{1}^{n}|u^{n}S^{n},X_{2}^{n}|v^{n}S^{n}%
}\left\{  \Pi_{X_{2}^{n}\left(  v^{n},S^{n}\right)  }\ \rho_{X_{1}^{n}\left(
u^{n},S^{n}\right)  ,X_{2}^{n}\left(  v^{n},S^{n}\right)  }\ \Pi_{X_{2}%
^{n}\left(  v^{n},S^{n}\right)  }\right\}  \right\}  \times\\
\mathcal{I}\left(  \left(  u^{\prime n},v^{\prime n},w^{\prime n}\right)
\in\mathcal{T}^{n}\right)  ,
\end{multline*}
where the inequality follows from dropping the indicator function
$\mathcal{I}\left(  \left(  u^{n},v^{n},w^{n}\right)  \in\mathcal{T}%
^{n}\right)  $. Considering that%
\begin{align*}
p\left(  u^{n},w^{n}|v^{n}\right)   &  =p\left(  u^{n}|v^{n}\right)
\delta_{w^{n},f^{n}\left(  u^{n}\right)  }\delta_{w^{n},g^{n}\left(
v^{n}\right)  }\\
&  \leq p\left(  u^{n}|v^{n}\right)  \delta_{w^{n},g^{n}\left(  v^{n}\right)
}%
\end{align*}
we have the upper bound:%
\begin{multline*}
\leq4\sum_{u^{\prime n},v^{\prime n},w^{\prime n}}\sum_{v^{n},w^{n}}p\left(
v^{n}\right)  \delta_{w^{n},g^{n}\left(  v^{n}\right)  }\times\\
\text{Tr}\left\{  \Gamma_{u^{\prime n},v^{\prime n},w^{\prime n}}%
\ \mathbb{E}_{S^{n}}\ \mathbb{E}_{X_{2}^{n}|v^{n}S^{n}}\left\{  \Pi_{X_{2}%
^{n}\left(  v^{n},S^{n}\right)  }\ \left(  \sum_{u^{n}}p\left(  u^{n}%
|v^{n}\right)  \mathbb{E}_{X_{1}^{n}|u^{n}S^{n}}\rho_{X_{1}^{n}\left(
u^{n},S^{n}\right)  ,X_{2}^{n}\left(  v^{n},S^{n}\right)  }\right)
\ \Pi_{X_{2}^{n}\left(  v^{n},S^{n}\right)  }\right\}  \right\}  \times\\
\mathcal{I}\left(  \left(  u^{\prime n},v^{\prime n},w^{\prime n}\right)
\in\mathcal{T}^{n}\right)  .
\end{multline*}%
\begin{multline*}
\leq4\sum_{u^{\prime n},v^{\prime n},w^{\prime n}}\sum_{v^{n},w^{n}}p\left(
v^{n}\right)  \delta_{w^{n},g^{n}\left(  v^{n}\right)  }\times\\
\text{Tr}\left\{
\begin{array}
[c]{c}%
\Pi\ \Pi_{S^{n}\left(  w^{\prime n}\right)  }\ \Pi_{X_{1}^{n}\left(  u^{\prime
n},S^{n}\right)  }\ \Pi_{X_{1}^{n}\left(  u^{\prime n},S^{n}\right)
,X_{2}^{n}\left(  v^{\prime n},S^{n}\right)  }\ \Pi_{X_{1}^{n}\left(
u^{\prime n},S^{n}\right)  }\ \Pi_{S^{n}\left(  w^{\prime n}\right)  }%
\ \Pi\ \times\\
\ \mathbb{E}_{S^{n}}\ \mathbb{E}_{X_{2}^{n}|v^{n}S^{n}}\left\{  \left(
\sum_{u^{n}}p\left(  u^{n}|v^{n}\right)  \mathbb{E}_{X_{1}^{n}|u^{n}S^{n}}%
\rho_{X_{1}^{n}\left(  u^{n},S^{n}\right)  ,X_{2}^{n}\left(  v^{n}%
,S^{n}\right)  }\right)  \right\}
\end{array}
\right\}  \times\\
\mathcal{I}\left(  \left(  u^{\prime n},v^{\prime n},w^{\prime n}\right)
\in\mathcal{T}^{n}\right)  .
\end{multline*}
where the second inequality follows from the fact that $\sum_{u^{n}}p\left(
u^{n}|v^{n}\right)  \mathbb{E}_{X_{1}^{n}|u^{n}S^{n}}\rho_{X_{1}^{n}\left(
u^{n},S^{n}\right)  ,X_{2}^{n}\left(  v^{n},S^{n}\right)  }$ and $\Pi
_{X_{2}^{n}\left(  v^{n},S^{n}\right)  }$ commute. Continuing, we have%
\begin{multline*}
=4\sum_{u^{\prime n},v^{\prime n},w^{\prime n}}\text{Tr}\left\{
\begin{array}
[c]{c}%
\Pi\ \Pi_{S^{n}\left(  w^{\prime n}\right)  }\ \Pi_{X_{1}^{n}\left(  u^{\prime
n},S^{n}\right)  }\ \Pi_{X_{1}^{n}\left(  u^{\prime n},S^{n}\right)
,X_{2}^{n}\left(  v^{\prime n},S^{n}\right)  }\ \Pi_{X_{1}^{n}\left(
u^{\prime n},S^{n}\right)  }\ \Pi_{S^{n}\left(  w^{\prime n}\right)  }%
\ \Pi\ \times\\
\ \left(  \sum_{v^{n},w^{n}}p\left(  v^{n}\right)  \delta_{w^{n},g^{n}\left(
v^{n}\right)  }\mathbb{E}_{S^{n}}\ \mathbb{E}_{X_{2}^{n}|v^{n}S^{n}}\left\{
\left(  \sum_{u^{n}}p\left(  u^{n}|v^{n}\right)  \mathbb{E}_{X_{1}^{n}%
|u^{n}S^{n}}\rho_{X_{1}^{n}\left(  u^{n},S^{n}\right)  ,X_{2}^{n}\left(
v^{n},S^{n}\right)  }\right)  \right\}  \right)
\end{array}
\right\}  \\
\times\mathcal{I}\left(  \left(  u^{\prime n},v^{\prime n},w^{\prime
n}\right)  \in\mathcal{T}^{n}\right)  .
\end{multline*}%
\begin{align*}
&  =4\sum_{u^{\prime n},v^{\prime n},w^{\prime n}}\text{Tr}\left\{  \Pi
_{S^{n}\left(  w^{\prime n}\right)  }\ \Pi_{X_{1}^{n}\left(  u^{\prime
n},S^{n}\right)  }\ \Pi_{X_{1}^{n}\left(  u^{\prime n},S^{n}\right)
,X_{2}^{n}\left(  v^{\prime n},S^{n}\right)  }\ \Pi_{X_{1}^{n}\left(
u^{\prime n},S^{n}\right)  }\ \Pi_{S^{n}\left(  w^{\prime n}\right)  }%
\ \Pi\ \overline{\rho}^{\otimes n}\ \Pi\right\}  \times\\
&  \ \ \ \ \ \ \ \mathcal{I}\left(  \left(  u^{\prime n},v^{\prime
n},w^{\prime n}\right)  \in\mathcal{T}^{n}\right)  \\
&  \leq4\ 2^{-n\left[  H\left(  B\right)  -\delta\right]  }\sum_{u^{\prime
n},v^{\prime n},w^{\prime n}}\text{Tr}\left\{  \Pi_{S^{n}\left(  w^{\prime
n}\right)  }\ \Pi_{X_{1}^{n}\left(  u^{\prime n},S^{n}\right)  }\ \Pi
_{X_{1}^{n}\left(  u^{\prime n},S^{n}\right)  ,X_{2}^{n}\left(  v^{\prime
n},S^{n}\right)  }\ \Pi_{X_{1}^{n}\left(  u^{\prime n},S^{n}\right)  }%
\ \Pi_{S^{n}\left(  w^{\prime n}\right)  }\ \Pi\ \right\}  \times\\
&  \ \ \ \ \ \ \ \mathcal{I}\left(  \left(  u^{\prime n},v^{\prime
n},w^{\prime n}\right)  \in\mathcal{T}^{n}\right)  \\
&  \leq4\ 2^{-n\left[  H\left(  B\right)  -\delta\right]  }\sum_{u^{\prime
n},v^{\prime n},w^{\prime n}}2^{n\left[  H\left(  B|X_{1}X_{2}\right)
+\delta\right]  }\ \mathcal{I}\left(  \left(  u^{\prime n},v^{\prime
n},w^{\prime n}\right)  \in\mathcal{T}^{n}\right)  \\
&  = 4\ 2^{-n\left[  H\left(  B\right)  -\delta\right]  }2^{n\left[
H\left(  B|X_{1}X_{2}\right)  +\delta\right]  }\sum_{u^{\prime n},v^{\prime
n},w^{\prime n}}\mathcal{I}\left(  \left(  u^{\prime n},v^{\prime n},w^{\prime
n}\right)  \in\mathcal{T}^{n}\right)  \\
&  \leq4\ 2^{-n\left[  H\left(  B\right)  -\delta\right]  }2^{n\left[
H\left(  B|X_{1}X_{2}\right)  +\delta\right]  }2^{n\left[  H\left(
UVW\right)  +\delta\right]  }\\
&  =4\ 2^{-n\left[  I\left(  X_{1}X_{2};B\right)  -H\left(  UVW\right)
-3\delta\right]  }.
\end{align*}
The first equality holds from the definition of $\overline{\rho}$
as the average output state
\[
	\overline{\rho}^{\otimes n} 
	= 
	\sum_{u^{n},v^{n}}p\left(  u^{n},v^{n}\right) 
	\mathbb{E}_{S^{n}}\ 
	\mathbb{E}_{X_{2}^{n}|v^{n}S^{n}}
	\mathbb{E}_{X_{1}^{n}|u^{n}S^{n}}
	\rho_{X_{1}^{n}\left(  u^{n},S^{n}\right)  ,X_{2}^{n}\left(v^{n},S^{n}\right)  } .
\]
The first inequality follows from \eqref{typ-prop-three}.
The second inequality follows because $\Pi_{X_{1}^{n}\left(  u^{\prime n},S^{n}\right)  }$,
$\Pi_{S^{n}\left(  w^{\prime n}\right)  }$ and $\Pi$ are all less than the identity 
and the property \eqref{eq:property-2-typical} of the projector $\Pi
_{X_{1}^{n}\left(  u^{\prime n},S^{n}\right)  ,X_{2}^{n}\left(  v^{\prime
n},S^{n}\right) }$.

\medskip

Thus, it is clear that as long as the information inequalities in the
statement of Theorem~\ref{thm:main-theorem} hold, then the expectation of the
error probability can be made arbitrarily small as $n$ becomes large. This
implies the existence of a particular code which accomplishes the desired
information processing task.
\end{proof}

\section{Conclusion}

\label{sec:conclusion}We have proved a quantum generalization of the Cover-El
Gamal-Salehi theorem, regarding joint source-channel encoding for transmission
of a classical correlated source over a multiple access channel. The theorem
should be useful in applications concerning communication over bosonic quantum
multiple access channels~\cite{YS05}. The main tools needed to prove this
theorem are an extension of those developed in the context of the quantum interference
channel \cite{S11a,FHSSW11}, combined with some further analysis using the \textquotedblleft
indicator function trick.\textquotedblright 

Future work might pursue extensions of Theorem~\ref{thm:main-theorem} to other
interesting scenarios:
\begin{enumerate}
\item Joint source-channel coding (JSCC) with extra side information \cite{RVS08}.
\item JSCC where the criterion is replaced by lossy transmission in the rate-distortion
sense~\cite{LMK10}.
\item The goal of the protocol is computation over the multiple access channel~\cite{NG07}.
\item JSCC in analog-digital hybrid coding scenarios \cite{MLK11}.
\end{enumerate}
Another very interesting pursuit would be the
\textquotedblleft fully quantum\textquotedblright\ generalization of this
problem, where the goal would be to transmit a correlated quantum source over
a quantum multiple access channel. The decoupling techniques from
Ref.~\cite{Horodecki:2007:107}\ should be useful in this context, and one
would expect a nice extension of the theorem from
Refs.~\cite{Horodecki:2007:107,YHD05MQAC}.

We acknowledge useful discussions with Pranab Sen. M.~M.~Wilde acknowledges
support from the Centre de Recherches Math\'{e}matiques. I.~Savov acknowledges
support from FQRNT and NSERC.

\appendix

\section{Typical Sequences and Typical Subspaces}

\label{sec:typ-review}Consider a density operator $\rho$ with the following
spectral decomposition:%
\[
\rho=\sum_{x}p_{X}\left(  x\right)  \left\vert x\right\rangle \left\langle
x\right\vert .
\]
The weakly typical subspace is defined as the span of all vectors such that
the sample entropy $\overline{H}\left(  x^{n}\right)  $ of their classical
label is close to the true entropy $H\left(  X\right)  $ of the distribution
$p_{X}\left(  x\right)  $ \cite{book2000mikeandike,W11}:%
\[
T_{\delta}^{X^{n}}\equiv\text{span}\left\{  \left\vert x^{n}\right\rangle
:\left\vert \overline{H}\left(  x^{n}\right)  -H\left(  X\right)  \right\vert
\leq\delta\right\}  ,
\]
where%
\begin{align*}
\overline{H}\left(  x^{n}\right)   &  \equiv-\frac{1}{n}\log\left(  p_{X^{n}%
}\left(  x^{n}\right)  \right)  ,\\
H\left(  X\right)   &  \equiv-\sum_{x}p_{X}\left(  x\right)  \log p_{X}\left(
x\right)  .
\end{align*}
The projector $\Pi_{\rho,\delta}^{n}$\ onto the typical subspace of $\rho$ is
defined as%
\[
\Pi_{\rho,\delta}^{n}\equiv\sum_{x^{n}\in T_{\delta}^{X^{n}}}\left\vert
x^{n}\right\rangle \left\langle x^{n}\right\vert ,
\]
where we have \textquotedblleft overloaded\textquotedblright\ the symbol
$T_{\delta}^{X^{n}}$ to refer also to the set of $\delta$-typical sequences:%
\[
T_{\delta}^{X^{n}}\equiv\left\{  x^{n}:\left\vert \overline{H}\left(
x^{n}\right)  -H\left(  X\right)  \right\vert \leq\delta\right\}  .
\]
The three important properties of the typical projector are as follows:%
\begin{align}
\text{Tr}\left\{  \Pi_{\rho,\delta}^{n}\rho^{\otimes n}\right\}   &
\geq1-\epsilon, \nonumber \\
\text{Tr}\left\{  \Pi_{\rho,\delta}^{n}\right\}   &  \leq2^{n\left[  H\left(
X\right)  +\delta\right]  }, \nonumber \\
2^{-n\left[  H\left(  X\right)  +\delta\right]  }\Pi_{\rho,\delta}^{n}  &
\leq\Pi_{\rho,\delta}^{n}\rho^{\otimes n}\Pi_{\rho,\delta}^{n}\leq2^{-n\left[
H\left(  X\right)  -\delta\right]  }\Pi_{\rho,\delta}^{n},  \label{typ-prop-three}
\end{align}
where the first property holds for arbitrary $\epsilon,\delta>0$ and
sufficiently large $n$. Consider an ensemble $\left\{  p_{X}\left(  x\right)
,\rho_{x}\right\}  _{x\in\mathcal{X}}$ of states. Suppose that each state
$\rho_{x}$ has the following spectral decomposition:%
\[
\rho_{x}=\sum_{y}p_{Y|X}\left(  y|x\right)  \left\vert y_{x}\right\rangle
\left\langle y_{x}\right\vert .
\]
Consider a density operator $\rho_{x^{n}}$ which is conditional on a classical
sequence $x^{n}\equiv x_{1}\cdots x_{n}$:%
\[
\rho_{x^{n}}\equiv\rho_{x_{1}}\otimes\cdots\otimes\rho_{x_{n}}.
\]
We define the weak conditionally typical subspace as the span of vectors
(conditional on the sequence $x^{n}$) such that the sample conditional entropy
$\overline{H}\left(  y^{n}|x^{n}\right)  $ of their classical labels is close
to the true conditional entropy $H\left(  Y|X\right)  $ of the distribution
$p_{Y|X}\left(  y|x\right)  p_{X}\left(  x\right)  $
\cite{book2000mikeandike,W11}:%
\[
T_{\delta}^{Y^{n}|x^{n}}\equiv\text{span}\left\{  \left\vert y_{x^{n}}%
^{n}\right\rangle :\left\vert \overline{H}\left(  y^{n}|x^{n}\right)
-H\left(  Y|X\right)  \right\vert \leq\delta\right\}  ,
\]
where%
\begin{align*}
\overline{H}\left(  y^{n}|x^{n}\right)   &  \equiv-\frac{1}{n}\log\left(
p_{Y^{n}|X^{n}}\left(  y^{n}|x^{n}\right)  \right)  ,\\
H\left(  Y|X\right)   &  \equiv-\sum_{x}p_{X}\left(  x\right)  \sum_{y}%
p_{Y|X}\left(  y|x\right)  \log p_{Y|X}\left(  y|x\right)  .
\end{align*}
The projector $\Pi_{\rho_{x^{n}},\delta}$ onto the weak conditionally typical
subspace of $\rho_{x^{n}}$ is as follows:%
\[
\Pi_{\rho_{x^{n}},\delta}\equiv\sum_{y^{n}\in T_{\delta}^{Y^{n}|x^{n}}%
}\left\vert y_{x^{n}}^{n}\right\rangle \left\langle y_{x^{n}}^{n}\right\vert
,
\]
where we have again overloaded the symbol $T_{\delta}^{Y^{n}|x^{n}}$ to refer
to the set of weak conditionally typical sequences:%
\[
T_{\delta}^{Y^{n}|x^{n}}\equiv\left\{  y^{n}:\left\vert \overline{H}\left(
y^{n}|x^{n}\right)  -H\left(  Y|X\right)  \right\vert \leq\delta\right\}  .
\]
The three important properties of the weak conditionally typical projector are
as follows:%
\begin{align}
\mathbb{E}_{X^{n}}\left\{  \text{Tr}\left\{  \Pi_{\rho_{X^{n}},\delta}%
\rho_{X^{n}}\right\}  \right\}   &  \geq1-\epsilon,\nonumber\\
\text{Tr}\left\{  \Pi_{\rho_{x^{n}},\delta}\right\}   &  \leq2^{n\left[
H\left(  Y|X\right)  +\delta\right]  },\label{eq:property-2-typical}\\
2^{-n\left[  H\left(  Y|X\right)  +\delta\right]  }\ \Pi_{\rho_{x^{n}}%
,\delta}  &  \leq\Pi_{\rho_{x^{n}},\delta}\ \rho_{x^{n}}\ \Pi_{\rho_{x^{n}%
},\delta}\leq2^{-n\left[  H\left(  Y|X\right)  -\delta\right]  }\ \Pi
_{\rho_{x^{n}},\delta}, \label{eq:property-3-typical}%
\end{align}
where the first property holds for arbitrary $\epsilon,\delta>0$ and
sufficiently large $n$, and the expectation is with respect to the
distribution $p_{X^{n}}\left(  x^{n}\right)  $.

An operator inequality that we make frequent use of is the \textquotedblleft
projector trick inequality\textquotedblright:%
\[
\Pi_{\rho_{x^{n}},\delta}\leq2^{n\left[  H\left(  Y|X\right)  +\delta\right]
}\rho_{x^{n}}.
\]
This follows from (\ref{eq:property-3-typical}) and the fact that $\Pi
_{\rho_{x^{n}},\delta}\ \rho_{x^{n}}\ \Pi_{\rho_{x^{n}},\delta}\leq\rho
_{x^{n}}$.

\section{Weak Joint Typicality and the Indicator Function Trick}

\label{sec:indicator-function-trick}Consider the joint random variable
$\left(  X,Y\right)  $ with distribution $p\left(  x,y\right)  $. We define
the weak jointly typical set $\mathcal{T}^{n}$\ as the set of all sequences
whose sample joint entropy is close to the true entropy and whose marginal
sample entropies are close to the true marginal entropies:%
\[
\mathcal{T}^{n}\equiv\left\{  \left(  x^{n},y^{n}\right)  :\left\vert
\overline{H}\left(  x^{n}\right)  -H\left(  X\right)  \right\vert \leq
\delta,\ \left\vert \overline{H}\left(  y^{n}\right)  -H\left(  Y\right)
\right\vert \leq\delta,\ \left\vert \overline{H}\left(  x^{n},y^{n}\right)
-H\left(  XY\right)  \right\vert \leq\delta\right\}  .
\]
This definition implies the following inequalities:%
\begin{align*}
\mathcal{I}\left(  \left(  x^{n},y^{n}\right)  \in\mathcal{T}^{n}\right)   &
\leq p\left(  x^{n}\right)  2^{n\left[  H\left(  X\right)  +\delta\right]
},\\
\mathcal{I}\left(  \left(  x^{n},y^{n}\right)  \in\mathcal{T}^{n}\right)   &
\leq p\left(  y^{n}\right)  2^{n\left[  H\left(  Y\right)  +\delta\right]
},\\
\mathcal{I}\left(  \left(  x^{n},y^{n}\right)  \in\mathcal{T}^{n}\right)   &
\leq p\left(  x^{n},y^{n}\right)  2^{n\left[  H\left(  XY\right)
+\delta\right]  }.
\end{align*}
Additionally, the restrictions on the marginal and joint sample entropies
imply that the following inequality holds for any $\left(  x^{n},y^{n}\right)
\in\mathcal{T}^{n}$:%
\[
\left\vert \overline{H}\left(  x^{n}|y^{n}\right)  -H\left(  X|Y\right)
\right\vert \leq\delta^{\prime},
\]
for some $\delta^{\prime}$ that is a function of $\delta$. In turn, this
inequality implies a different \textquotedblleft indicator function
trick\textquotedblright:%
\[
\mathcal{I}\left(  \left(  x^{n},y^{n}\right)  \in\mathcal{T}^{n}\right)  \leq
p\left(  x^{n}|y^{n}\right)  2^{n\left[  H\left(  X|Y\right)  +\delta^{\prime
}\right]  }.
\]

\section{Gentle Operator Lemma}

\label{sec:useful-lemmas}

\begin{lemma}
[Gentle Operator Lemma for Ensembles \cite{itit1999winter,ON07,W11}%
]\label{lem:gentle-operator} Given an ensemble $\left\{  p_{X}\left(
x\right)  ,\rho_{x}\right\}  $ with expected density operator $\rho\equiv
\sum_{x}p_{X}\left(  x\right)  \rho_{x}$, suppose that an operator $\Lambda$
such that $I\geq\Lambda\geq0$ succeeds with high probability on the state
$\rho$:%
\[
\mathrm{Tr}\left\{  \Lambda\rho\right\}  \geq1-\epsilon.
\]
Then the subnormalized state $\sqrt{\Lambda}\rho_{x}\sqrt{\Lambda}$ is close
in expected trace distance to the original state $\rho_{x}$:%
\[
\mathbb{E}_{X}\left\{  \left\Vert \sqrt{\Lambda}\rho_{X}\sqrt{\Lambda}%
-\rho_{X}\right\Vert _{1}\right\}  \leq2\sqrt{\epsilon}.
\]

\end{lemma}

\section{Marginal and Conditional Distributions formed from the Source
Distribution}

In this appendix, we detail several properties of the source distribution
$p\left(  u,v,w\right)  $, under the assumption that $W$ is the common part of
$\left(  U,V\right)  \sim p\left(  u,v\right)  $ (as described in the main
text). First, we demonstrate the following equality:%
\[
\sum_{w}p\left(  u,w|v\right)  =p\left(  u|v\right)
\]
Consider that%
\begin{align}
\sum_{w}p\left(  u,w|v\right)   &  =\sum_{w}\frac{p\left(  u,w,v\right)
}{p\left(  v\right)  }\nonumber\\
&  =\sum_{w}\frac{p\left(  u,v\right)  \delta_{w,f\left(  u\right)  }%
\delta_{w,g\left(  v\right)  }}{p\left(  v\right)  }\nonumber\\
&  =\frac{p\left(  u,v\right)  \sum_{w}\delta_{w,f\left(  u\right)  }%
\delta_{w,g\left(  v\right)  }}{p\left(  v\right)  }\nonumber\\
&  =\frac{p\left(  u,v\right)  }{p\left(  v\right)  }\nonumber\\
&  =p\left(  u|v\right)  \label{eq:source-1}%
\end{align}
Similarly, we have%
\begin{equation}
\sum_{w}p\left(  v,w|u\right)  =p\left(  u|v\right)  .\label{eq:source-2}%
\end{equation}

We would like to show that%
\[
p\left(  v|u,w\right)  =p\left(  v|u\right)  ,
\]
which seems reasonable since knowledge of $u$ and $w$ should be the same as
knowledge of $u$ alone (given that $w$ is computed from $u$). We have%
\begin{align*}
p\left(  v|u,w\right)   &  =\frac{p\left(  u,v,w\right)  }{p\left(
u,w\right)  }\\
&  =\frac{p\left(  w|u,v\right)  p\left(  u,v\right)  }{p\left(  w|u\right)
p\left(  u\right)  }\\
&  =\frac{\delta_{w,f\left(  u\right)  }\delta_{w,g\left(  v\right)  }p\left(
v|u\right)  }{\delta_{w,f\left(  u\right)  }}%
\end{align*}
This distribution is only defined whenever $w=f\left(  u\right)  $.

\enlargethispage{\baselineskip}

\bibliographystyle{plain}
\bibliography{Ref}

\end{document}